\documentclass[11pt]{article}
\usepackage{amssymb}
\usepackage{amsmath}
\usepackage{graphicx}
\usepackage{color}

\def\beq{\begin{eqnarray}}    
\def\eeq{\end{eqnarray}}      

\newcommand{\CC}{\Lambda}
\newcommand{\rLo}{\rho_{\CC}^0}
\newcommand{\rL}{\rho_{\CC}}
\newcommand{\rmr}{\rho_m}

\newcommand{\LQCD}{\Lambda_{\rm QCD}}
\newcommand{\ba}{\begin{eqnarray}}
\newcommand{\ea}{\end{eqnarray}}
\newcommand{\brr}{\begin{array}}
\newcommand{\err}{\end{array}}
\newcommand{\bc}{\begin{center}}
\newcommand{\ec}{\end{center}}

\newcommand{\be}{\begin{equation}}
\newcommand{\ee}{\end{equation}}

\newcommand{\OMB}{\Omega_B}

\newcommand{\ODM}{\Omega_{\rm DM}}

\newcommand{\alem}{\alpha_{\rm em}}
\newcommand{\qL}{q_{\Lambda}}

\begin{document}

\hyphenation{cos-mo-lo-gists un-na-tu-ral-ly ne-cessa-ry dri-ving
par-ti-cu-lar-ly a-na-ly-sis mo-del mo-dels ex-ten-ding e-xam-ples
ho-we-ver res-pec-ti-ve-ly}

\begin{center}

{\LARGE \textit{Running vacuum in the Universe and the time variation\\ of the fundamental constants of Nature}} \vskip 2mm

 \vskip 8mm

 \vskip 8mm

\textbf{Harald Fritzsch\,$^{a,b}$,  Joan Sol\`a\,\,$^{a,c,d}$}\footnote{E-mail: sola@fqa.ub.edu \vskip2mm}

\vskip0.25cm

$^{a}$ Institute for Advanced Study, Nanyang Technological University, Singapore

$^{b}$ Physik-Department, Universit\"at M\"unchen, D-80333 Munich, Germany

$^{c}$  Departament de  F\'\i sica Qu\`antica i Astrof\'\i sica,  Universitat
de Barcelona, \\ Av. Diagonal 647,
 E-08028 Barcelona, Catalonia, Spain\\

 $^{d}$ Institute of Cosmos Sciences, Universitat de Barcelona (ICCUB) \\ Av. Diagonal 647,
 E-08028 Barcelona, Catalonia, Spain

 \vspace{0.5cm}

{\bf Rafael C. Nunes}

Dept. de F\'\i sica, Universidade Federal de Juiz de Fora, 36036-330, \\ Juiz de Fora, MG, Brazil.

\end{center}
\vskip 15mm

\begin{abstract}

 We compute the time variation of the fundamental constants (such as the ratio of the proton mass to the electron mass, the strong coupling constant, the fine structure constant and Newton's constant) within the context of the so-called \textit{running vacuum models} (RVM's) of the cosmic evolution. Recently, compelling evidence has been provided that these models are able to fit the main cosmological data (SNIa+BAO+H(z)+LSS+BBN+CMB) significantly better than the concordance $\CC$CDM model. Specifically, the vacuum parameters of the RVM (i.e. those responsible for the dynamics of the vacuum energy) prove to be nonzero at a confidence level $\gtrsim3\sigma$.
 Here we use such remarkable status of the RVM's to make definite predictions on the cosmic time variation of the fundamental constants. It turns out that the predicted variations are close to the present observational limits. Furthermore,  we find  that the time evolution of the dark matter particle masses should be crucially involved in the total mass variation of our Universe. A positive measurement of this kind of effects could be interpreted as strong support to the ``micro and macro connection'' (viz. the dynamical feedback between the evolution of the cosmological parameters and the time variation of the fundamental constants of the microscopic world), previously proposed by two of us (HF and JS).


\vskip0.2cm

\end{abstract}

\newpage

\section{Introduction}
The possibility that the so-called ``constants" of Nature (such as the particle masses and the couplings associated to their interactions) are actually not constants, but time evolving quantities following the slow pace of the current cosmological evolution, has been investigated in the literature since long time ago\,\cite{FritzschBook}.
The history of these investigations traces back to early ideas in the thirties on the possibility of a time evolving gravitational constant $G$ by Milne\,\cite{Milne1935} and the suggestion by Dirac of the large number hypothesis \cite{Dirac1937}, which led him also to propose in 1937 the time evolution of $G$. The same year Jordan speculated that the fine structure constant $\alpha_{\rm em}$ together with $G$ could be both space and time dependent\,\cite{Jordan1937} -- see also \cite{JordanBook}. This, more field-theoretically oriented, point of view was retaken later on by Fierz\,\cite{Fierz1956} and, finally, in the sixties, $G$ was formally associated to the existence of a dynamical scalar field $\phi\sim 1/G$ coupled to the curvature.  Such was the famous framework originally proposed by Brans and Dicke (``BD'' for short) \cite{BD,DickeReviews57and61}, which was the first historical attempt to extend General Relativity (GR) to accommodate variations in the Newtonian coupling $G$. To avoid conflict with the weak equivalence principle, the variability of $\phi$ in the BD-theory was originally restricted to a possible time evolution only, i.e. $\phi(t)$ with no spatial dependence \,\cite{Damour2012}.  A generalization of the BD approach in the seventies\,\,\cite{Wagoner1970} subsequently led to a wide variety of scalar-tensor theories, which are still profusely discussed in the current literature. Furthermore, the subject of the time variation of the fine structure constant (cf. \cite{Gamow} for other early proposals) is particularly prolific. Numerous studies have been undertaken in our days from different perspectives, sometimes pointing to positive observational evidence\,\cite{alphat1} but often disputed by alternative observations\,\cite{alphat2} --  see e.g. the reviews \cite{Fritzsch2009,Barrow2010,Uzan2011,Calmet2015}.

The possibility that the particle masses could also drift with the cosmic evolution has also become a favorite target of the modern astrophysical observations. Thus e.g. the dimensionless proton-to-electron mass ratio, $\mu \equiv m_p/m_e$,
has been carefully monitored using quasar absorption lines.  Claims on significant time variation
$\dot{\mu}/\mu$ at $\sim$ 4$\sigma$ c.l. are available in the literature\,\cite{Reinhold2006}, although still unconfirmed by other observations\,\cite{King2008}.
Future high precision quantum optic experiments in the lab are also planned to test the possible time variation of these observables,
and they will most likely be competitive\,\cite{FritzschBook}. Clearly, the time and space variation of the fundamental constants is a very active
field of theoretical and experimental research that may eventually provide interesting surprises in the near future
\cite{Preface}.

From a more modern perspective the dark energy (DE) theories of the cosmological evolution also predict the cosmic time variation of the fundamental constants. These include the string-dilaton models, Kaluza-Klein theories, chameleon models etc., in which the underlying dynamical fields are either massless or nearly massless -- cf.  e.g. \cite{TimeConstants} for some early attempts and \cite{Uzan2011} for a review of more recent proposals and the observational situation. The possibility that dark matter theories could also impinge on the time variation of fundamental constants has also been put forward\,\cite{Flambaum2015}. Many other proposals are available in the literature, see e.g. \,\cite{SpecialIssueMPLA} for more contemporary developments.

In this paper, we would like to focus our attention on a specific class of DE models leading to a time variation of the fundamental constants\,\cite{FritzschSola2012}. These models are particularly interesting since it has recently been shown that they prove capable of fitting the main cosmological data in a fully competitive way with the concordance $\CC$CDM model\,\cite{planck2015}. These are the so-called running vacuum models (RVM's) of the cosmological evolution, see e.g.\,\cite{MG14} for a recent summarized discussion and \cite{JSPRev2013,SolaGomezReview,JSPReview2016} for a more extensive review and a comprehensive list of references. In these models there are no ultralight scalar fields and the time variation of the fundamental constants is effectively triggered via quantum effects induced by the cosmological renormalization group, whose flow is naturally set up by the expansion rate $H$, cf.\,\cite{JSPRev2013} and references therein. The framework is characterized by a dynamical vacuum energy density, $\rL$, which is a power series of $H$ and its time derivatives. For the current Universe, it suffices to consider $\rL=\rL(H,\dot{H})$ up to linear terms in $\dot{H}$ and quadratic in $H$\,\cite{SolaGomezReview,JSPReview2016,GomSolaBas15,GomSola2015,Elahe2015}. However,  extensions with higher powers have also been considered for describing inflation, see \,\cite{SolaGomezReview,GRF2015} and\cite{LBS1,LBS2,LBS3} for different kind of scenarios, including anomaly-induced inflation\,\cite{Fossil07}. Because of the dynamical nature of the vacuum in these models, a natural feedback occurs between the cosmological parameters and the fundamental constants of the microscopic world, such as the particle masses and coupling constants. Remarkably, recent studies have shown that the class of RVM's can provide an excellent fit to the main cosmological data (SNIa+BAO+H(z)+LSS+BBN+CMB) which is highly competitive with that of the $\CC$CDM -- see most particularly \,\cite{SolaGomCruzApJL,SolaGomCruzApJ,PRL2017,SolaGomCruzMPLA}, and previous studies such as \cite{GomSolaBas15,GomSola2015,BPS09,GSBP11} and references therein. Therefore, there is plenty of motivation for further investigating these
running vacuum models. In this paper, building upon the aforementioned works which single out the especial status of the RVM's, we wish to  estimate the possible variation of the fundamental constants triggered by the dynamical interplay between the evolution of the vacuum energy density $\rL=\rL(H)$ and the concomitant anomalous conservation law of matter (which may lead to time dependent particle masses) and/or the time evolution of the gravitational coupling $G(H)$.  Such feedback between the ultralarge scales of cosmology and the minute scales of the subatomic physics is what two of us have called elsewhere ``the micro and macro connection''\,\cite{FritzschSola2015,FritzschSola2012}. Recently it has been shown that it could also lead to a possible explanation for the origin of the Higgs potential in the gravitational framework\,\cite{SolaKarim2017}.

This paper is organized as follows. After a preliminary historical discussion in the introduction, in sect. 2 we remind of the possibility of the cosmic time variation of the cosmological parameters in the context of GR without committing to any particular model. In sect. 3 we focus on  the running vacuum models (RVM's)  as a particularly appealing framework where to realize the time evolution of the fundamental constants. In sect. 4 we consider in detail the specific prediction of the RVM's concerning the time variation of the particle masses and couplings. In sect. 5 we briefly discuss alternative dynamical vacuum models. Finally, in sect. 6 we present our conclusions.

\section{Cosmological models with time evolving parameters}
\label{sec:models}

We wish to explore the possibility that the fundamental `constants' or parameters ${\cal P}$ of the subatomic world (such as masses and couplings) are actually slowly varying with (cosmic) time $t$ and whether this feature might be related to the cosmic evolution of the fundamental constants of gravitation. If so the cosmic time evolution of ${\cal P}$ should be typically proportional to the rate of change of the scale factor of the cosmic expansion, i.e.  $\dot{\cal P}/{\cal P}\propto\dot{a}/a\equiv H$ (the Hubble rate). From this point of view we should expect (at least in linear approximation) that the fractional cosmic drift rate of ${\cal P}(t)$ near our time is proportional to $H_0=1.0227\,h\times 10^{-10}\,{\rm
yr}^{-1}\,$, with $h\simeq 0.67$, and hence very small in a natural way. Specifically $\dot{\cal P}/{\cal P}\sim H_0\,\Delta{\cal P}/{\cal P}\lesssim \left(\Delta{\cal P}/{\cal P}\right)\,10^{-10}$yr$^{-1}$. From the existing bounds on the known particles, typically we expect that $\Delta{\cal P}/{\cal P}$ should vary between a few parts per million (ppm) to at most one part in a hundred thousand over a cosmological span of time, depending on the specific parameter (such as couplings, masses etc: ${\cal P}= G, m_i,\alpha_{\rm em}, \alpha_s, \Lambda_{\rm QCD}$...)  However, the attributes of the dark matter (DM) particles (e.g. their masses and couplings) could vary faster.
The possible cosmic time evolution of these parameters can equivalently
be described as a redshift dependence:
\begin{equation}\label{eq:dotffp}
\frac{\dot{\cal P}}{\cal P}=-(1+z)\,H(z)\,\frac{{\cal P}'(z)}{{\cal P}(z)}\,,
\end{equation}
where $z=(1-a)/a$ is the cosmic redshift, $a(t)$ is the scale factor
(normalized to $a(t_0)=1$ at present) and ${\cal P}'(z)=d{\cal P}/dz$ . It will prove useful to present most of our results in terms of the cosmological redshift.

The above micro-and-macro-connection scenario\,\cite{FritzschSola2012,FritzschSola2015} can be implemented from Einstein's field equations in the presence of a time-evolving cosmological constant, $\CC(t)$. In fact, without violating the Cosmological Principle in the context of the Friedmann-Lema\^\i tre-Robertson-Walker (FLRW) universe, nothing prevents $\CC(t)$ and/or  $G=G(t)$ from being functions of the cosmic time\,\footnote{The mere phenomenological possibility of a time-varying $\Lambda$ has been considered by many authors from different perspectives, see e.g. \cite{CCtPheno} for some examples and \cite{Overduin} for a review and more references.}. The field equations can be written $ G_{\mu\nu} = 8 \pi G \tilde{T}_{\mu\nu}$, where $G_{\mu\nu} = R_{\mu\nu} - 1/2 g_{\mu\nu} R$ and $\tilde{T}_{\mu\nu} = T_{\mu\nu} + g_{\mu\nu} \rho_{\Lambda}$ stand respectively for
the Einstein tensor and the full energy-momentum tensor, in which $T_{\mu\nu}$ is the ordinary part (involving only the matter fields) and $g_{\mu\nu} \rho_{\Lambda}$ carries the time-evolving vacuum energy density $\rho_{\Lambda}(t) = \Lambda (t)/ 8 \pi G(t)$.

The field equations for a variable gravitational ``constant'' and a dynamical vacuum energy density in the FLRW metric in
flat space are derived in the standard way and are formally similar to the case with $G$ and $\Lambda$ constants. The corresponding generalization of the Friedmann equation reads

\begin{equation}
\label{Friedmann}
 3 H^2 = 8 \pi G(t) [\rho_m(t) + \rho_{\Lambda}(t)],
\end{equation}
where $\rho_m = \rho_B + \rho_{DM}$  represents the contribution of non-relativistic matter (baryons + dark matter) and  $\rho_{\Lambda}(t)$ stands for the aforementioned dynamical vacuum energy density.

An important consequence of the covariance of GR is the Bianchi identity involving the Einstein tensor: $\nabla^{\mu}G_{\mu\nu} = 0$.
Via the field equations, it implies $\nabla^{\mu}(G \tilde{T}_{\mu\nu})=0$. Using the FLRW metric and considering that the matter of the current Universe is made of pressureless dust, we find the explicit form of the Bianchi identity is: $\dot{G}(\rho_m + \rho_{\Lambda}) + G(\dot{\rho}_m + \dot{\rho_{\Lambda}}) + 3GH \rho_m = 0$. This equation plays the role of a generalized local conservation law. Using Eq.\,(\ref{eq:dotffp}) it can be conveniently rewritten in terms of the redshift variable as follows:
\begin{equation}\label{GeneralizedConservation}
\frac{G'(z)}{G(z)}\,\left[\rmr(z)+\rL(z)\right]+{\rho}'_m(z)+\rho'_{\CC}(z)=\frac{3\rmr(z)}{1+z}\,.
\end{equation}
The above expression relates, in a fully general form, the evolution of the matter energy density and the vacuum energy density in the presence of a dynamical $G$.  For $G=$const.  and $\rL=$const. (i.e. within the context of the $\CC$CDM) Eq.\,(\ref{GeneralizedConservation}) integrates trivially and renders the canonical conservation law of non-relativistic matter:
\begin{equation}\label{LCDMconservation}
\rho_m(z)=\rho_m^0\,(1+z)^3\,,
\end{equation}
where $\rho_m^0$ is the current matter energy density.
Equations \eqref{Friedmann} and \eqref{GeneralizedConservation} cannot be solved beyond the $\CC$CDM unless some model or an ansatz is provided e.g. on the evolution of the vacuum energy density $\rL$. In the next section we adopt the proposal for $\rL$ associated to the running vacuum models (RVM's). In such case the solution for the matter energy density will no longer be in general of the canonical form (\ref{LCDMconservation}) and this, as we shall see, can be interpreted either as an anomalous matter conservation law or as a time evolution of the particle masses. Together with the cosmic evolution of $G$, we can see that this scenario may well lead us to a concrete realization of the time variation of the fundamental constants that can be consistent with GR.

\section{Running vacuum in the expanding Universe}
\label{sec:RVM}

Even though we know that the standard matter conservation law (\ref{LCDMconservation}) must be essentially correct, we wish to explore the possibility that small corrections could perhaps be accommodated. Let us first proceed phenomenologically and later on we will adduce some theoretical motivations supporting the obtained results.
Suppose that the standard dilution law for matter in an expanding universe, $\rho_m \propto a^{-3}$, receives a small correction:  $\rho_m \propto a^{-3(1-\nu_m)}$, with $|\nu_m|\ll 1$ a dimensionless parameter. In terms of the redshift, we have
\begin{equation}
\label{MatterEvolution}
  \rho_m = \rho_{m}^0 (1+z)^{3(1- \nu_m)},
\end{equation}
where for $\nu_m =0$ we retrieve of course the standard form (\ref{LCDMconservation}).  This possibility was first proposed and tested in the literature in Ref.\,\cite{Cristina2003}, and later on it was  addressed by other authors -- see e.g.\,\cite{WangMeng2004,AlcanizLima2005}. However, here we wish to reinterpret the anomalous conservation law (\ref{MatterEvolution}) in a different way\,\cite{FritzschSola2012}. Rather than assuming that the presence of a nonvanishing $\nu_m$ is related to an anomalous nonconservation of the number density of particle masses, we assume that it parameterizes the nonconservation of the particle masses themselves. In other words, we suppose that while there is a normal dilution of the particle number density with the expansion for all the particle species, i.e. $n_i=n_i^0\,a^{-3}=n_i^0(1+z)^3$, the corresponding mass values $m_i$ are not conserved throughout the cosmic expansion:
\begin{equation}\label{eq:mz}
m_i(z)=m_i^0\,(1+z)^{-3\nu_i}\,,
\end{equation}
where $m_i^0$ are the current values ($z=0$) of the particle masses and the various $\nu_i$ are the corresponding anomaly indices for the non-conservation of each species. This was the point of view adopted in\,\cite{FritzschSola2012}. Notice that $\rho_{m_i}$ in Eq.\,(\ref {MatterEvolution}), for the ith-species, is indeed equal to $n_i(z) m_i(z)$, with  $ m_i(z)$ given by (\ref{eq:mz}) and $\rho_{m_i}^0=n_i^0\,m_i^0$.  In the remaining of this section it will not be necessary to distinguish among the different values of $\nu_i$ for each particle species, but we shall return to this point in the next section.

\subsection{Running $G$ and running $\rL$}
\label{sec:RGLambda}

Let us assume that $G$ is also a slowly varying cosmic variable. More specifically, we shall suppose that it varies logarithmically with the Hubble function as follows:
\begin{equation}
\label{Gevolution}
 G(H) = \frac{G_0}{1 + \nu_G \ln \frac{H^2}{H^2_0} }\ ,
\end{equation}
where $\nu_G$ is another small dimensionless parameter ($|\nu_G|\ll 1$), different from $\nu_m$ in general.  A theoretical motivation for this expression within QFT in curved spacetime can be found in\,\cite{Fossil07} -- see also \cite{JSPRev2013}. There is also a practical reason for this form, as we will see in a moment.

The above assumptions stand us in good stead to find out what is the corresponding dynamical evolution law for the vacuum energy density  within GR. As indicated, we assume that the parameters are small since  the ensuing variation of the fundamental constants must be small and the model cannot depart significantly from the standard cosmology. Having this in mind and substituting the expressions (\ref{MatterEvolution}) and (\ref{Gevolution}) in the generalized conservation law (\ref{GeneralizedConservation}), as well as
using the Friedmann equation (\ref{Friedmann}), we can analytically integrate the evolution of $\rL$ as a function of the redshift. The final result reads as follows:
\begin{equation}
\label{VacuumEvolution}
 \rho_{\Lambda} = \rho_{\Lambda}^0 + \frac{\nu_m + \nu_G}{1 -\nu_m -\nu_G}\, \rho_{m}^0\, [(1+z)^{3(1-\nu_m)} -1]\,.
\end{equation}
Here $\rL(z=0)=\rL^0$ is the current value of the vacuum energy density, and of course this value would stay constant for the entire cosmic history (as in the $\CC$CDM) if both $\nu_m$ and $\nu_G$ were zero. Therefore $\nu_m + \nu_G$ can be interpreted as an effective coefficient for the running of $\rho_{\Lambda}$. The obtained result (\ref{VacuumEvolution}) shows that the dynamical character of the vacuum energy can indeed be in interplay with both the dynamics of the gravitational coupling and the non-conservation of matter, in a manner perfectly compatible with the GR field equations.  Amusingly, we note that if $\nu_m=-\nu_G$ the vacuum energy density would remain constant as in the $\CC$CDM, although the model would not quite behave as the standard cosmological model because the anomalous matter conservation law (\ref{MatterEvolution}) and the logarithmically evolving Newtonian coupling (\ref{Gevolution}) both stay in force.  In any case the departure from the standard cosmology in all cases will be small enough provided $|\nu_m|\ll 1$ and $|\nu_G|\ll 1$.

Defining the normalized Hubble function with respect to the current value, $E=H/H_0$, explicit computation from Friedmann's equation, using (\ref{MatterEvolution}) and (\ref{VacuumEvolution}), provides the following result: %
\begin{equation}\label{eq:HzRVM}
E^2(z)=\frac{G}{G_0}\left\{1+\frac{\Omega_m^0}{1-\nu_m-\nu_G}\left[(1+z)^{3(1-\nu_m)}-1\right]\right\}\,,
\end{equation}
where $G$ is given in our case by (\ref{Gevolution}). Notice that for $\nu_m=0$ (matter conservation) and $\nu_G=0$ ($G=$const.) the Hubble function (\ref{eq:HzRVM}) boils down to the $\CC$CDM, as should be expected.

We may now understand why the forms (\ref{MatterEvolution}) and (\ref{Gevolution}), leading to (\ref{VacuumEvolution}), are particularly interesting. The obtained result for the vacuum energy density can be motivated from the integration of the renormalization group (RG) flow associated to the Hubble expansion, within the context of the running vacuum model (RVM) -- see \cite{JSPRev2013,SolaGomezReview,GomSolaBas15,GomSola2015} and references therein. Let us note that the leading RG effects up to order $\sim H^4$ take on the form of the power series\,\cite{SolaGomezReview}
 \be
\rL(H)
=a_0+a_1\,\dot{H}+a_2\,H^2+a_3\,\dot{H}^2+a_4\,H^4+a_5\,\dot{H}\,H^2+...\,,
\label{GRVE} \ee
where the coefficients $a_i$ have different dimensionalities in
natural units. Specifically, $a_0$ has dimension $4$ since this is
the dimension of $\rL$; $a_1$ and $a_2$ have both dimension $2$; and,
finally, $a_3$, $a_4$ and $a_5$ are dimensionless. Notice that only the terms with an even number of derivatives of the scale factor are present in the above expression since the vacuum energy density is part of the effective action of QFT in curved spacetime and therefore must preserve general covariance. The ${\cal O}(H^4)$ terms can be important for the inflationary stage \,\cite{SolaGomezReview,GRF2015,LBS1,LBS2,LBS3}, but for the post-inflationary epoch and in particular for the current Universe they can be neglected.  We are thus left with the ${\cal O}(H^2)$ terms only, namely $H^2$ and $\dot{H}$. For simplicity we will focus on the former since the inclusion of the latter will not affect the main discussion in this study concerning the variation of the fundamental constants\,\footnote{The general cosmological solution involving  both the $H^2$ and $\dot{H}^2$ terms in the vacuum energy density has been discussed in detail in \,\cite{SolaGomezReview,GomSolaBas15,GomSola2015,SolaGomCruzApJL,SolaGomCruzApJ}}. Therefore we concentrate here on the simplest form of the RVM density, which we can rewrite after an appropriate redefinition of $a_0$ and $a_2$ as follows:
\begin{eqnarray}\label{eq:ModelAgeneral}
\rL(H)&=&\rLo+\frac{3\nu}{8\pi}\,M_P^2\, (H^2-H_0^2)\,,
\end{eqnarray}
where we have arranged that for $H=H_0$ we recover the current value of the vacuum energy density: $\rL(H_0)=\rLo$. Here $M_P=1/\sqrt{G_0}$ is the Planck mass and $\nu$ is a small dimensionless coefficient  which characterizes the dynamical evolution of $\rL(H)$; in fact $\nu$ plays the role of the $\beta$-function coefficients of the renormalization group equation (RGE) for the running vacuum\,\footnote{For a concrete estimate of $\nu$ in QFT in curved spacetime, see\,\cite{Fossil07}. One finds the general form $\nu\sim\sum_iM_i^2/M_P^2$, where $M_i$ are the masses of fermions and bosons in the loops within a typical Grand Unified scenario. See also \cite{JSPRev2013}.}.
In particular, for $\nu=0$ we recover the $\CC$CDM case. Notice that  $M_P^2\,H^2$ is of order $\rLo$ and hence for $\nu\neq0$ the term $\sim\nu\,M_P^2\,H^2$ represents a small correction (for $|\nu|\ll1$) to the constant value $\rLo$  and endows  $\rL$  of a mild dynamical behavior which can be favorable to observations. Indeed, by fitting the parameter $\nu$ to the overall cosmological data one finds an improved fit as compared to the $\CC$CDM, provided $|\nu|\sim 10^{-3}$ -- for details, see\,\cite{SolaGomCruzApJL,SolaGomCruzApJ,GomSolaBas15,GomSola2015}.

The connection between (\ref{eq:ModelAgeneral}) and (\ref{VacuumEvolution}) can now be elucidated as follows\,\cite{SolaGomezReview,GomSolaBas15}. To start with, take $G=$const. as this simplifies the structure of the generalized conservation law (\ref{GeneralizedConservation}). Inserting (\ref{MatterEvolution}) in that law we can solve for $\rL(z)$  and we find the expression (\ref{VacuumEvolution}) with $\nu_G=0$. Knowing the matter and vacuum densities, Friedmann's equation immediately provides $H$ as a function of the redshift and we arrive at Eq.\, (\ref{eq:HzRVM}) (with $G=G_0$).  Combining this expression for $H(z)$ with that of $\rL(z)$ we find (\ref{eq:ModelAgeneral}).

One can also relate the running vacuum law (\ref{eq:ModelAgeneral}) with a running gravitational coupling of the form (\ref{Gevolution}) as follows. Assume now that the standard local matter conservation law (\ref{LCDMconservation})  holds good, i.e. $\nu_m=0$ in (\ref{MatterEvolution}). In this way  Eq.\,(\ref{GeneralizedConservation}) boils down to $(dG/G)(\rho_m+\rho_{\CC})+d\rho_\CC=0$. The latter can be solved using (\ref{eq:ModelAgeneral}) with $\nu=\nu_G$ and noting also that $\rho_m+\rho_{\CC}=3H^2/(8\pi G)$ by virtue of Friedmann's equation (\ref{Friedmann}). The result is a simple differential equation for $G(H)$:
\begin{equation}\label{eq:diffGH}
\frac{dG}{G^2}=-\nu_G\,M_P^2\frac{dH^2}{H^2}=-\frac{\nu_G}{G_0}\frac{dH^2}{H^2}\,.
\end{equation}
Integrating it with the boundary condition $G(H_0)=G_0=1/M_P^2$,  the final result is indeed Eq.\,(\ref{Gevolution}). In other words, if matter is conserved the running vacuum (\ref{eq:ModelAgeneral})  leads to a logarithmic evolution of the gravitational coupling of the form (\ref{Gevolution}). These two evolution laws take therefore the precise dynamical forms needed to fulfill the Bianchi identity of the field equations when matter is conserved.

In the previous considerations we have assumed either  $\nu_G=0$ or $\nu_m=0$. But we can also determine the running of $\rL$ as a function of $H$ when $\nu_m$ and $\nu_G$ are both nonvanishing.  From (\ref{VacuumEvolution}) and (\ref{eq:HzRVM}) we find:
 \be
\rL(H)
=\rho_c^0(1-\Omega_m^0)+\frac{3\nu}{8\pi G_0}\left(\frac{G_0}{G}H^2-H_0^2\right)\,,
\label{rLH2} \ee
with $\nu=\nu_m+\nu_G$. Notice that $\rho_c^0=3H_0^2/(8\pi G_0)$ is the current critical density and hence $\rho_c^0(1-\Omega_m^0)=\rho_{\Lambda}^0$. However, since $G_0/G=1+{\cal O}(\nu_G)$, see Eq.(\ref{Gevolution}), we find that Eq.\,(\ref{rLH2}) boils down to Eq.\,(\ref{eq:ModelAgeneral}) at first order in the small parameters $(\nu_m,\nu_G)$. Recall that in all cases these parameters are assumed to satisfy $|\nu_m,\nu_G|\ll1$. It follows that  Eq.\,(\ref{eq:ModelAgeneral}) does correctly describe the running of the vacuum energy density in terms of $H$ in all possible cases within this framework, i.e. even when matter is non-conserved and at the same time there is a running of the gravitational coupling. In other words, the running of $\rL$ as a function of $H$ in (\ref{eq:ModelAgeneral}) is controlled in all cases by $\nu=\nu_m+\nu_G$, irrespective of whether one or none of these parameters is zero.

\subsection{Free parameters}
\label{sec:Parameters}

Let us summarize the situation with the free parameters in the class of RVM's considered here.  The basic parameters are $(\nu_m,\nu_G)$, which are associated to the anomalous matter conservation law (\ref{MatterEvolution}) and the time evolution of the gravitational coupling, Eq.\,(\ref{Gevolution}). Given these two parameters the Bianchi identity (\ref{GeneralizedConservation}) determines the evolution of the vacuum energy density, Eq.\,(\ref{VacuumEvolution}), through  $\nu=\nu_m+\nu_G$. However, as we have mentioned, there is no reason to expect that all of the particle masses should have the same anomaly index, and hence we expect that the anomaly mass index $\nu_m$ can be expressed in terms of the different particle indices $\nu_i$ defined in Eq.\,(\ref{eq:mz}). For example, baryons can have the index $\nu_B$ (assumed the same for all of them), but it could be different from the index for DM particles, X, which we call $\nu_X$. The relation between the overall anomaly index $\nu_m$ and the specific indices $(\nu_B,\nu_X)$ will be discussed in sect. 4.

Furthermore, as indicated, if it turns that  $a_1\neq0$ in Eq.\,(\ref{GRVE}), then an additional parameter is still possible for the running of the vacuum energy in the present universe, but we shall not take it into account since it is not necessary to illustrate the possible existence of the basic effects under study. In actual fact, in all phenomenological considerations we will assume $a_1=0$ together with one of the following two possibilities: either i) $\nu_m\neq 0$ with $\nu_G=0$, or ii) $\nu_G\neq0$ with $\nu_m=0$. This will suffice to parametrize the time variation of the fundamental constants that we are considering here. In such context the Bianchi identity enforces the value of $\nu$ (the parameter that controls the running of the vacuum energy density in (\ref{eq:ModelAgeneral})) to be either  $\nu_m$ or $\nu_G$, depending on wether we assume either that $G$ is fixed and the matter has some anomaly conservation law, or that the matter is strictly conserved and $G$ has some evolution, but not both situations at the same time.  While $\nu$ could perfectly receive simultaneous contributions from both kinds of effects --in the above mentioned form $\nu=\nu_m+\nu_G$ -- at the moment it is not possible to individually disentangle them phenomenologically. Thus, in our numerical evaluations we will always assume the separate situations in which either $\nu=\nu_m$ or $\nu=\nu_G$.

From the foregoing considerations we see that the RVM's offer several possibilities for the time variation of the fundamental ``constants'', all of them being connected through the evolution laws (\ref{MatterEvolution}), (\ref{Gevolution}), (\ref{VacuumEvolution}) and (\ref{eq:ModelAgeneral}), in which there is a built-in  principle for exchanging energy  between matter, vacuum and the gravitational coupling in different combinations that are compatible with GR.

\section{Time evolution of the fundamental constants in the RVM}
\label{sec:Fundamental Constants}

In this section we wish to evaluate the specific impact of the running vacuum models (RVM's) on the time evolution of the fundamental constants of the standard model (SM) of particle physics and the fundamental constants of cosmology. Basically we will assess the predicted variation of the particle masses, most conspicuously the proton mass (through the proton-to-electron mass ratio $\mu\equiv m_p/m_e$), the QCD scale $\LQCD$, the QED fine structure constant $\alpha_{\rm em}$, the corresponding QCD coupling $\alpha_s$, the gravitational constant $G$ and the cosmological constant $\CC$. We will also consider the possible implications from Grand Unified Theories (GUT's).

\subsection{Time variation of masses and couplings in the SM}\label{sec:RunMass}

The framework outlined in sect.\,\ref{sec:RVM} suggests that basic quantities of the standard model (SM) of strong and electroweak interactions, such as the quark masses, the proton mass and the quantum chromodynamics (QCD)
scale parameter, $\Lambda_{QCD}$, might not be conserved in the course of the cosmological evolution \cite{FritzschSola2012,FritzschSola2015}.
Let us take, for example, the proton mass, which is given as follows:
$m_p = c_{QCD} \Lambda_{QCD} + c_u m_u + c_d m_d + c_s m_s + c_{em} \Lambda_{QCD}$, where $m_{u,d.s}$ are the quark masses and the last term represents the electromagnetic (em) contribution.  Obviously the leading term is the first one, which is due to the strong binding energy of QCD. Thus, the nucleon mass can be expressed to within very good approximation as $m_p \simeq c_{QCD} \LQCD \simeq 938 MeV$, in which
$c_{QCD}$ is a non-perturbative coefficient. The masses of the light quarks $m_u$,
$m_d$ and $m_s$ also contribute to the proton mass, although by less than 10$\%$ and can therefore be neglected for this purpose. It follows that time (or cosmic redshift) variations of the proton mass are essentially equivalent to time (redshift) variations of the QCD scale parameter:
\begin{equation}\label{eq:mpQCD}
\frac{\dot{m}_p}{m_p}\simeq \frac{\dot{\Lambda}_{\rm QCD}}{\LQCD}=-(1+z)\,H(z)\,\frac{m_p'(z)}{m_p}\,,
\end{equation}
where in the last equality we have used Eq\,(\ref{eq:dotffp}).
On the other hand, the QCD scale parameter is related to the strong coupling constant $\alpha_s = g^2_s/4 \pi$ as follows (at 1-loop order):
\begin{equation}\label{eq:alphasLQCD}
 \alpha_s(\mu_R) = \frac{4 \pi}{(11-2 n_f/3) \ln( \mu^2_R/\Lambda^2_{QCD})},
\end{equation}
$\mu_R$ being the renormalization point, $n_f$ the number of quark flavors and $\Lambda_{QCD}$  = 217 $\pm$ 25 MeV the measured value of the QCD scale paramete. However, if there is a crosstalk between the micro and macro world as suggested in the previous section, we expect that when we embed QCD in the context of a FLRW expanding universe the value of the proton mass, and hence of $\Lambda_{QCD}$, need not remain constant anymore. The possible change of $\Lambda_{QCD}$ should be of course relatively small, and from the above considerations we envisage that its value may evolve with the rate of expansion of the universe, i.e. $\Lambda_{QCD} = \Lambda_{QCD}(H)$. In such case the strong coupling constant $\alpha_s$ becomes a function not only of the conventional renormalization scale  $\mu_R$  but also of the cosmic scale $\mu_c\equiv H$. Since $H=H(z)$ is a function of the cosmological redshift, we can write  $\alpha_s=\alpha_s(\mu_R, z)$.
From Eq.\,(\ref{eq:alphasLQCD}) we find that the relative variations of the two QCD quantities with the Hubble rate are related (at one-loop) in the following manner:
\begin{equation}
\label{alpha_s}
\frac{1}{\alpha_s} \frac{d \alpha_s(\mu_R,z)}{dz} = \frac{1}{\ln(\mu_R/\LQCD(z))} \Big[ \frac{1}{\LQCD(z)}\frac{d \LQCD(z)}{dz} \Big]\,.
\end{equation}
If the QCD coupling constant $\alpha_s$ or the QCD scale parameter $\Lambda_{QCD}$ undergo a small cosmological shift, the nucleon masses and the masses of the atomic nuclei would also change along with $\Lambda_{QCD}$. The cosmic dependence of the
strong coupling $\alpha_s(\mu_R, z)$ can be generalized to the electroweak couplings of the SM, including the fine structure constant in QED, $\alpha_{\rm em}$, except that there is no electroweak equivalent for the $\Lambda_{QCD}$ scale. As a consequence there is no obvious connection of the variation of the particle masses with the variation of the electroweak coupling. This new sort of time variation would be possible only if the vacuum expectation value of the Higgs would be itself time-dependent, but we shall not address this option here. There is however an alternative link based on the framework of Grand Unified Theories, which will be explored in sect.\,\ref{sect:alphaem}.


\begin{figure}[t]
\begin{center}
\label{Hz}
\includegraphics[width=4in, height=3in]{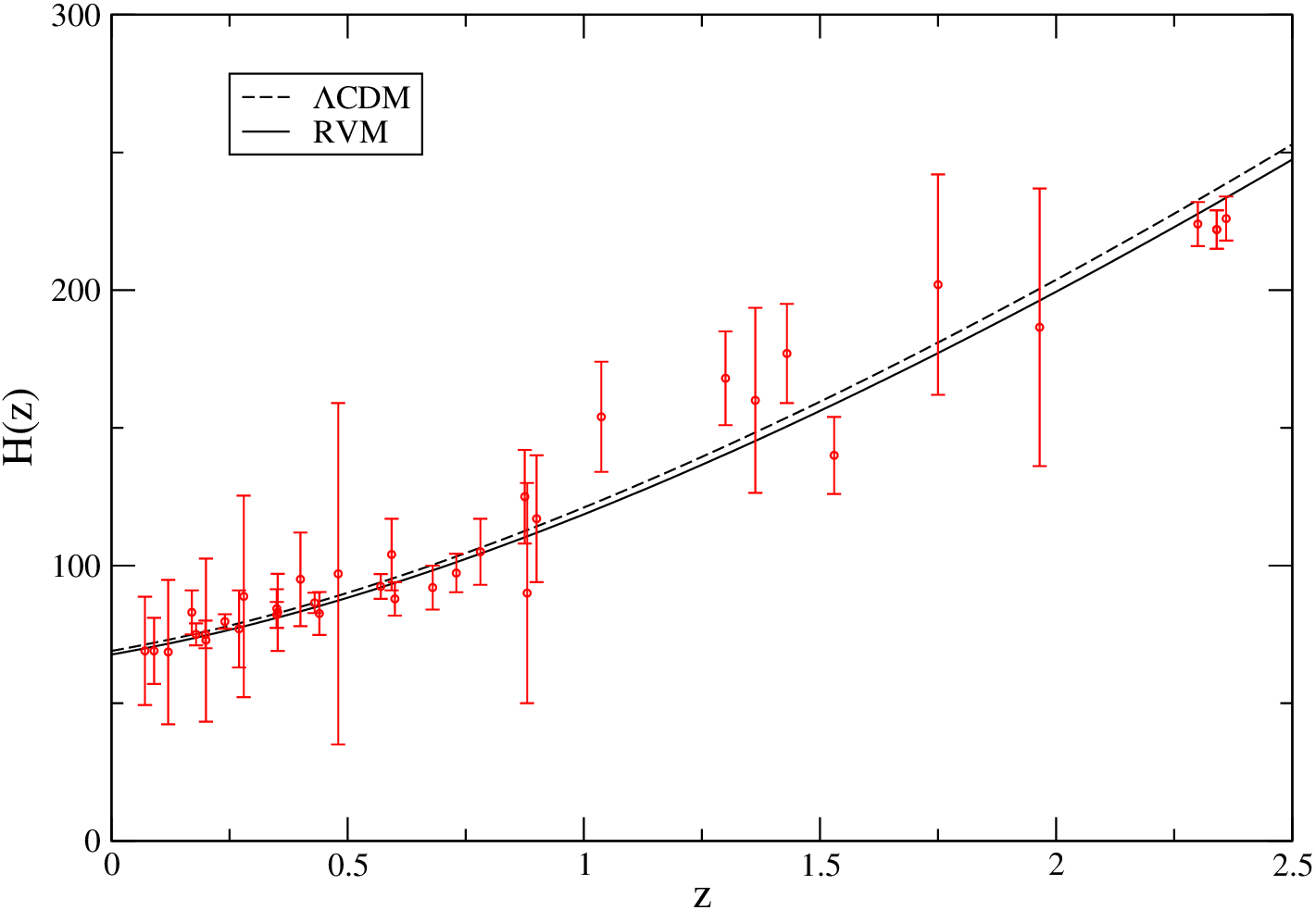}
\caption{Data on the Hubble rate $H(z)$ (in Km/sec/Mpc) at different redshifts versus the theoretical Hubble function of the running vacuum model (RVM) in the $G=$const. case (solid line) and the $\CC$CDM (dashed line), see text. The value of the RVM vacuum parameter is fixed at  $\nu =0.001$, as this is the order of magnitude obtained in the joint likelihood fit to the overall SNIa+BAO+H(z)+LSS+BBN+CMB data performed in\,\cite{SolaGomCruzApJL,SolaGomCruzApJ,PRL2017} (see references therein). }
\end{center}
\end{figure}

\begin{figure}[t]
\begin{center}
\label{Hznu1}
\includegraphics[width=5in, height=3in]{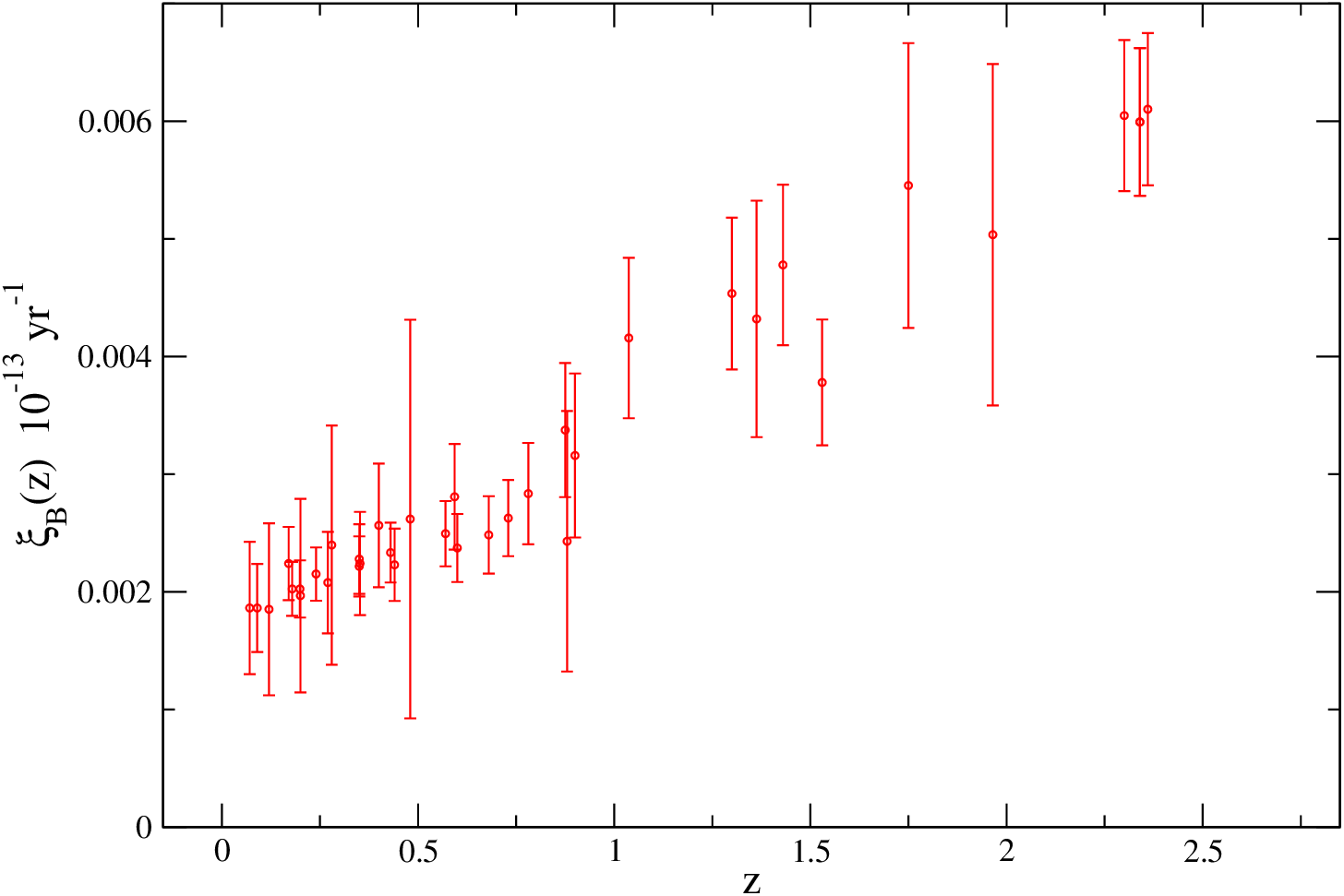}
\caption{The specific contribution from baryons
to the total mass drift rate. We plot the dimensionless function $\xi_B$ from (\ref{indices}) for $\nu_B=10^{-5}$ using the $H(z)$ data of Fig. 1 (see text).}
\end{center}
\end{figure}


\begin{figure}[t]
\begin{center}
\label{Hznu2}
\includegraphics[width=4in, height=3in]{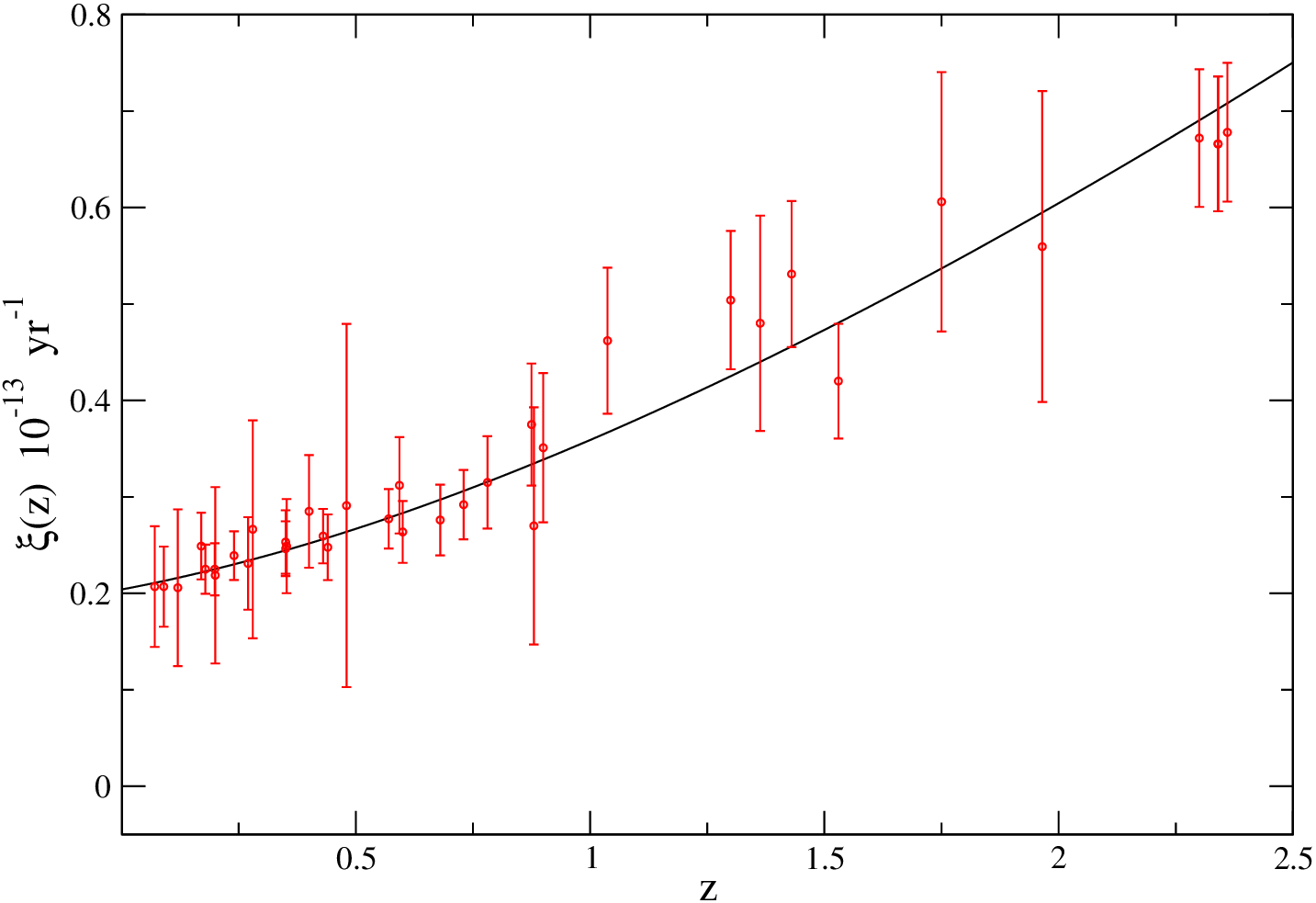}
\caption{The total mass drift rate $\xi(z)$, given by Eq.\,(\ref{xifunction}) as predicted by the RVM,
as a function of the redshift and within the same conditions as in the previous figure. Here $\nu_X=10^{-3}$, which essentially saturates the fitted value of $\nu_m$ (recall that $\nu_B\ll\nu_m$, see the text).}
\end{center}
\end{figure}



\begin{figure}[t]
\begin{center}
\label{Hznu3}
\includegraphics[width=4in, height=3in]{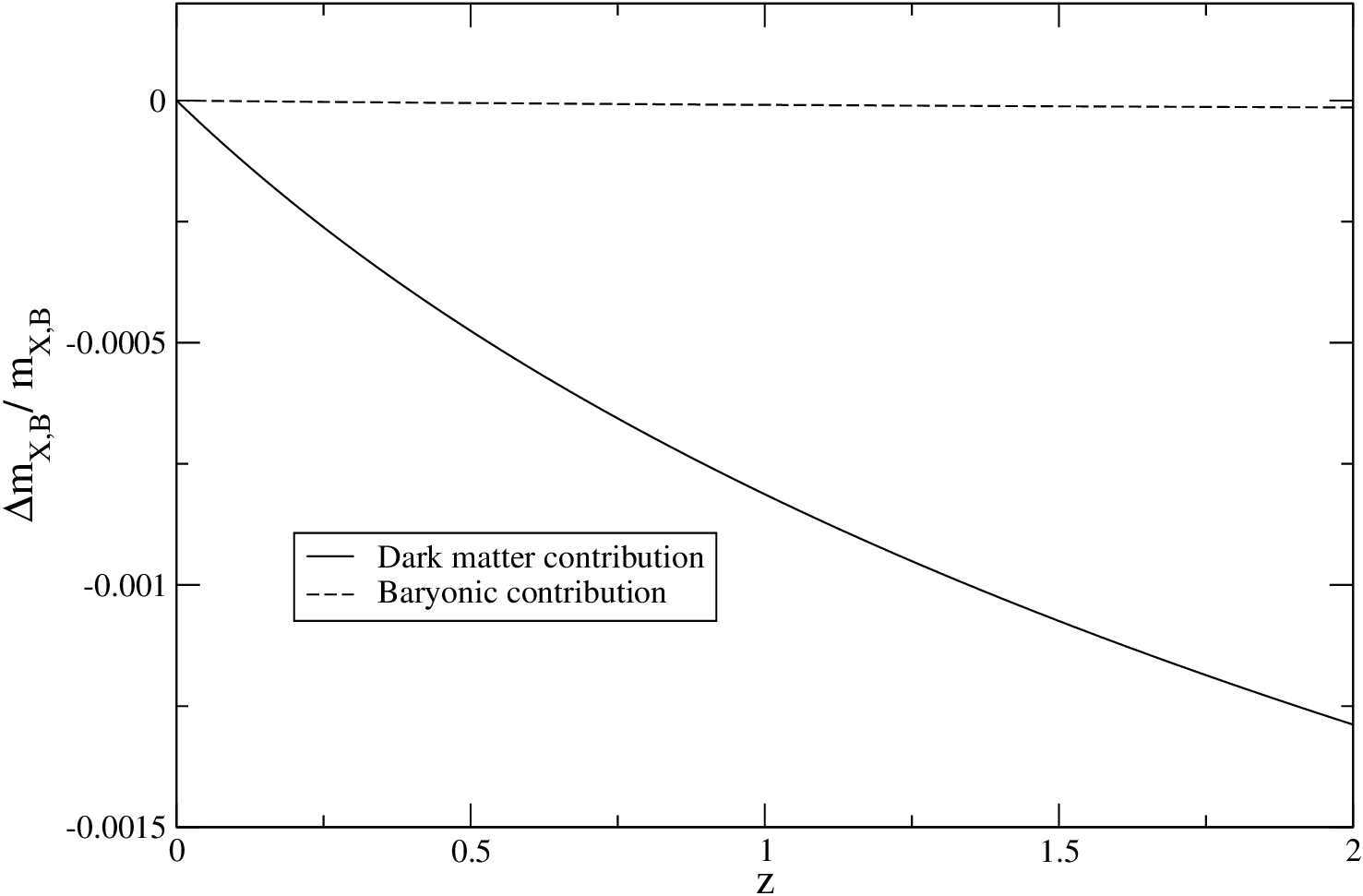}
\caption{The mild time evolution of the baryon masses (dashed line) versus the more substantial evolution of the DM particle masses (solid line). As before, $\nu_B=10^{-5}$ and  $\nu_X=10^{-3}$. }
\end{center}
\end{figure}


In this paper we attribute the cosmic variation of the particle masses to the energy exchange with the cosmic vacuum according to the RVM framework outlined in the previous section, in which a possible additional variation of the gravitational constant may also concur.
In order to estimate quantitatively these effects within the RVM, we take as a basis the numerical fit estimates obtained in \cite{SolaGomCruzApJL,SolaGomCruzApJ,GomSolaBas15,GomSola2015} using the known data on SNIa+BAO+H(z)+LSS+BBN+CMB -- see also \cite{JSPReview2016} for a review.  Among these observational sources (which involve several hundreds of data points indicated in these references) we use 36 data points on the Hubble rate $H(z)$ at different reshifts in the range $0 < z \leq 2.36$, as compiled in \cite{Hdata,RatraFarooq2013},
out of which 26 data points are inferred from the differential age method, whereas 10 correspond to measures obtained from the
baryonic acoustic oscillation method (cf. Fig.\,1). These data will play a significant role in our aim to constrain cosmological parameters because they are obtained from model-independent direct observations.  In particular, we use this compilation for investigating a possible temporal evolution of the particle masses, both for baryons and dark matter.

In Fig. 1 we plot the above mentioned data points $H(z_i)\ (i=1,2,...,36)$ and at the same time we superimpose the Hubble functions $H(z)$ for both the $\CC$CDM model (dashed line) and the RVM (solid one). The difference between the two is small, of course, because the RVM Hubble function (\ref{eq:HzRVM}) differs only mildly from the standard one owing to the parameters $\nu_m$, $\nu_G$ being small in absolute value. However, the small differences are perfectly visible in Fig.\,1 and are sufficient to improve significantly the overall fit to the SNIa+BAO+H(z)+LSS+BBN+CMB data points. According to \cite{GomSolaBas15}, the data show a preference (at the level of $\gtrsim3\sigma$) for a dynamical vacuum of the form (\ref{eq:ModelAgeneral}) rather than the rigid vacuum ($\nu=0$) of the $\CC$CDM case, for which $\rL=$const. The obtained fit values of the vacuum parameter $\nu$  stay in the ballpark of $10^{-3}$\cite{SolaGomCruzApJL,SolaGomCruzApJ,JSPReview2016}, and therefore we can use this order of magnitude determination as a characteristic input in our estimate of the variation of the fundamental constants. The predicted mass drift rates are indicated in Figs. 2-3, and in what follows we explain their origin.

To estimate the variation of the particle masses within the RVM, let us start by noting that the matter density of the universe can be approximated as: $\rho_m \simeq n_p m_p + n_n m_n + n_{X} m_{X}$,
where we neglect the leptonic contribution and the relativistic component (photons and neutrinos)\,\cite{FritzschSola2012}.
Here $n_p , n_n , n_{X}$ ($m_p , m_n , m_{X}$) are the number densities (and corresponding masses) of protons, neutrons and dark matter (DM) particles $X$, respectively. Assuming that the mass non-conservation law in eq. (\ref{MatterEvolution}) is to be attributed to the change of the mass of the particles -- confer. Eq. (\ref{eq:mz}) -- the relative total time variation of the mass density associated to such mass anomaly can be estimated as follows:
\begin{equation}\label{eq:reltimeMdensityUniv1}
\frac{\delta\dot{\rho}_m}{\rho_m}\simeq
\frac{n_p\,\dot{m}_p+n_n\,\dot{m}_n+n_X\,\dot{m}_X}{n_X\,m_X}\,\left(1-\frac{\Omega_B}{\Omega_{DM}}\right)\,,
\end{equation}
where $\delta\dot{\rho}_m$ is obtained by differentiating $\rho_m $ with respect to time and subtracting the ordinary (i.e. fixed mass) time dilution of the number densities. In the above equation, $\Omega_B$ and $\Omega_{DM}$ represent as usual the fractional density of baryons and DM particles with respect to the critical density, respectively. Of course the total $\Omega_m$ is the sum $\Omega_B+\Omega_{DM}$ and in the numerical analysis we take as a current value $\Omega_m=0.30$, according to the global fits obtained from the RVM models\,\cite{SolaGomCruzApJL,SolaGomCruzApJ,PRL2017}. The approximation made on  the r.h.s. of (\ref{eq:reltimeMdensityUniv1}) uses the fact that  $\rho_B/\rho_{DM}=\Omega_B/\Omega_{DM}={\cal O}(10^{-1})$, with $\rho_B=n_pm_p+n_nm_n$ and $\rho_{DM}=n_Xm_X\equiv\rho_X$ the density of DM particles.  Equation (\ref{eq:reltimeMdensityUniv1}) can be further expanded as follows. Let us take  $m_n = m_p \equiv m_B$ so that $\rho_B=(n_p+n_n)\, m_B$, and assume $\dot{m}_n = \dot{m}_p\equiv\dot{m}_B$. Since $n_n/n_p$ is of order $10\%$ after the primordial nucleosynthesis and $\Omega_B/\Omega_{DM}$ is also of order $10\%$,  we may neglect the product of these two terms or any higher power of them. In this way we are led to
\begin{equation}\label{eq:reltimeMdensityUniv2}
\frac{\delta\dot{\rho}_m}{\rho_m}=\frac{n_p\,\dot{m}_B}{n_X\,m_X}\,\left(1+\frac{n_n}{n_p}-\frac{\OMB}{\ODM}\right)+
\frac{\dot{m}_X}{m_X}\left(1-\frac{\OMB}{\ODM}\right)\,.
\end{equation}
Note that the prefactor in the first term of the \textit{r.h.s.} can be written
\begin{equation}\label{eq:prefactor}
\frac{n_p\,\dot{m}_B}{n_X\,m_X}=\frac{(\rho_B-n_nm_B) \dot{m}_B}{\rho_X m_B}=\frac{\OMB}{\ODM}\,\frac{\dot{m}_B}{m_B}\left(1-\frac{n_n/n_p}{1+n_n/n_p}\right)\simeq
\frac{\OMB}{\ODM}\,\frac{\dot{m}_B}{m_B}\left(1-\frac{n_n}{n_p}\right)\,.
\end{equation}
Inserting \eqref{eq:prefactor} in \eqref{eq:reltimeMdensityUniv2} we see that the leading power of $n_p/n_n$ appears at second order and hence can be neglected. The final result therefore reads:
\begin{equation}
\label{mass1}
 \frac{\delta \dot\rho_m}{\rho_m} \simeq \Big(1- \frac{\Omega_B}{\Omega_{DM}} \Big) \Big(\frac{\Omega_B}{\Omega_{DM}} \frac{\dot{m}_B}{m_B} + \frac{\dot{m}_{X}}{m_{X}} \Big)\,.
\end{equation}
At this point we are ready to relate the RVM prediction of the total mass drift throughout the cosmic expansion with the individual mass variations of baryons and dark matter.  With the help of Eq. (\ref{MatterEvolution}), the total fractional mass density variation with time can  be written  $\delta \dot\rho_m/ \rho_m \simeq 3 \nu_m H$, in linear approximation of the small parameter $\nu_m$ and for moderate values of the redshift\,\cite{FritzschSola2012}. Inserting this expression on the \textit{l.h.s} of Eq.\,(\ref{mass1}) we can rewrite it in the following convenient way:
\begin{equation}
\label{mass2}
\frac{\nu_m }{1 - \Omega_B/\Omega_{DM}}=\frac{\Omega_B}{\Omega_{DM}} \nu_B + \nu_{X}\,,
\end{equation}
where we have introduced the anomaly indices $\nu_B$ and $\nu_{X}$ for the evolution of the baryon and DM masses. They define the corresponding mass drift rates for baryons and DM particles:
\begin{equation}
\label{indices}
\xi_B(t)\equiv\frac{\dot{m_B}}{m_B} = 3 \nu_B H, \,\,\,\,\,\,\,\,\,\,\,\,\,\,\,\,\,  \xi_X(t)\equiv\frac{\dot{m}_{X}}{m_{X}}= 3 \nu_{X}H\,.
\end{equation}
The total drift rate from the time variation of the masses of all heavy and stable particles in the Universe (baryons + dark matter) reads
\begin{equation}
\label{xifunction}
\xi(t) = \xi_B(t)+\xi_X(t) = 3H (\nu_B + \nu_{X}).
\end{equation}
The drift rates are of course functions of time and redshift. The corresponding relation with the variation of a particular mass $m_i$, baryon or DM, within a cosmological span of time $\Delta t\sim H^{-1}$ can be a complicated function of time, but it is usually approximated in a linear way, i.e. one assumes that on average the time variation of the mass was the same during the time interval $\Delta t$. In this way we can write
\begin{equation}\label{eq:variationmi}
\frac{\dot{m}_i}{m_i}\simeq \frac{\Delta m_i}{m_i\,\Delta t}\simeq \frac{\Delta{m_i}}{m_i}\,H \ \ \ \rightarrow\ \ \  \frac{\Delta{m_i}}{m_i}\simeq 3\nu_i\,.
\end{equation}
Thus, the anomaly indices $\nu_i$ encode the typical mass variation of a given particle species (baryons or DM particles) in a cosmological span of time.

Since, as mentioned, $\Omega_B/\Omega_{DM}\simeq 0.1$, we can neglect the square of this quantity  and rewrite (\ref{mass2}) in the more compact form
\begin{equation}
\label{mass3}
\nu_m =\frac{\Omega_B}{\Omega_{DM}} \left(\nu_B-\nu_{X}\right) + \nu_{X}\,.
\end{equation}
This is the promised relation between the anomaly mass index in Eq.\,(\ref{MatterEvolution}) and the specific baryonic and DM anomaly indices. Let us note that the presence of these anomaly indices for matter non-conservation can affect different aspects of the cosmic history, such as the precise moment of matter-radiation equality or the details of the growth of structure formation. These effects have been evaluated in different works, see e.g. \cite{RVMgrowth,PRL2017,GomSola2015}, and can be important in the future when more precise observational data will be available. For the present work, it suffices to estimate the order of magnitude of the anomaly indices in (\ref{mass3}) in order to transfer the possible impact on the time variation of the fundamental constants. For this reason we used here only the order of magnitude values of the fitting results obtained in\,\cite{SolaGomCruzApJL,SolaGomCruzApJ,PRL2017}.
As far as the lepton index $\nu_L$ is concerned, it cannot have any significant effect in the above equation (\ref{mass3}) since it is suppressed by the small lepton mass rate in the Universe as compared to baryons. See, however, Eq.\,(\ref{eq:Deltamuovermu}) below for other effects from $\nu_L$ that might be not so negligible.

Equation (\ref{mass3}) can actually be checked experimentally, for $\nu_m$ is related to the running of the vacuum energy density (e.g. $\nu_m=\nu$ when $\nu_G=0$, as explained in sect. \ref{sec:RVM}) and $\nu$ can be fitted from cosmological observations based on SNIa+BAO+H(z)+LSS+BBN+CMB data\,\cite{SolaGomCruzApJL,SolaGomCruzApJ,JSPReview2016}, and it is found to be of order $10^{-4}-10^{-3}$, whereas $\nu_B$ can be determined from astrophysical and lab experiments usually aimed at determining the time evolution of the ratio $\mu=m_p/m_e$\,\cite{Uzan2011,SpecialIssueMPLA}. Thus, if the equation (\ref{mass2}) -- or equivalently (\ref{mass3}) -- must be fulfilled, we can indeed check if the DM part $\nu_{X}$ (which, of course, cannot be measured individually) plays a significant role in it.

\begin{figure}
\label{G1}
\begin{center}
\includegraphics[width=3in, height=3in]{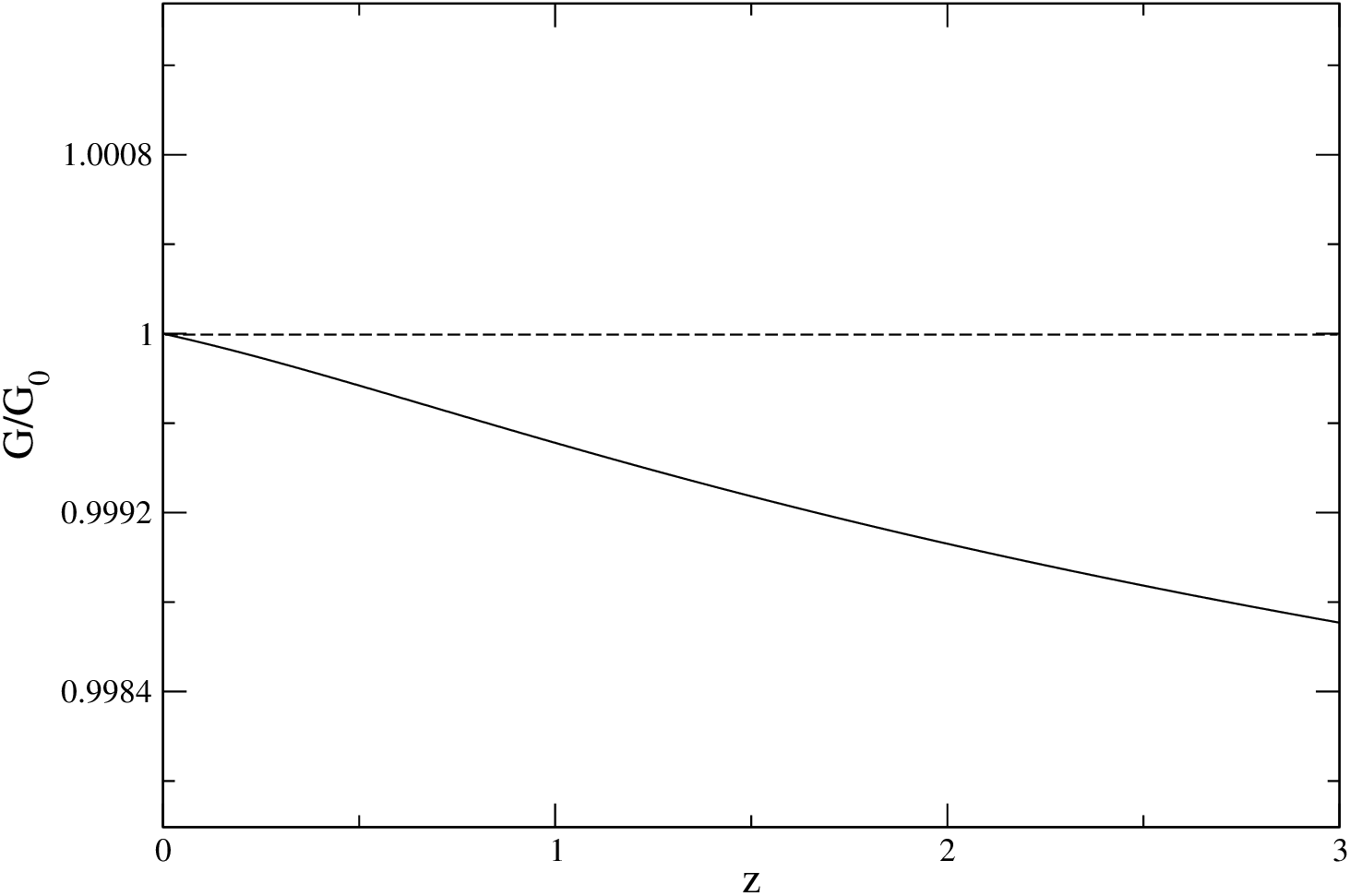}\hspace{1cm}
\includegraphics[width=3in, height=3in]{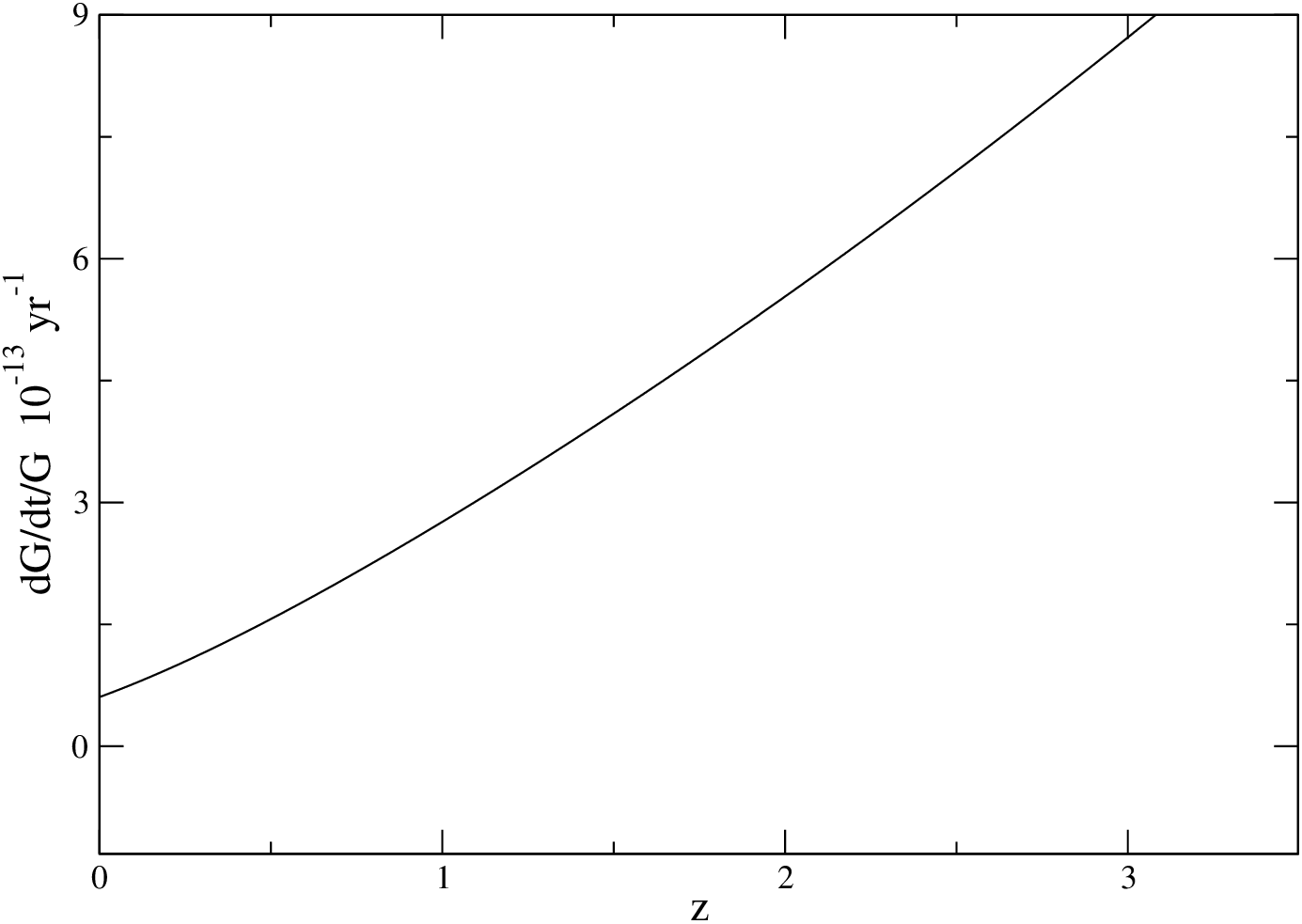}
\caption{Left panel: Evolution of $G(z)/G_0$ based on Eq. (\ref{Gevolution}) at leading order in $\nu_G$.
We use $\nu_G = 0.001$ from the fit of the RVM to the overall cosmological data\,\cite{SolaGomCruzApJL,SolaGomCruzApJ,PRL2017}.
Right panel: As in the left panel, but now we plot the cosmic drift rate ${\dot{G}}/{G}$ of the gravitational coupling as a function of the
redshift, according to the RVM expression (\ref{eq:dotGsmaex}).}
\end{center}
\end{figure}


Observationally one finds from a rich variety of experimental situations both from astrophysical observations and direct lab measurements\,\cite{Uzan2011,Calmet2015,Ferreira} (most of them compatible with a null result) that $\Delta\mu/\mu$ is at most in the ballpark of ${\cal O}(1-10)$ parts per million (ppm). Let us note that
\begin{equation}\label{eq:Deltamuovermu}
\frac{\Delta\mu}{\mu}=\frac{\Delta m_p}{m_p}-\frac{\Delta m_e}{m_e}=3(\nu_B-\nu_L)\,,
\end{equation}
where we have used Eq.\,(\ref{eq:variationmi}). The index $\nu_B$ was applied to the proton as the only stable baryon, whereas $\nu_L$ corresponds to the electron as the only stable lepton. It is usually assumed that $\nu_B\gg\nu_L$ and then $\Delta\mu/\mu\simeq\Delta m_p/m_p$. In this case the aforementioned limit on $\Delta\mu/{\mu}$ would imply $\nu_B\sim 10^{-5}$ at most. However, a more symmetric option (which cannot be ruled out at present) is that the two indices $\nu_B$ and $\nu_L$ can be close to each other. In such case both could be of order $10^{-4}$ and very similar; this case would still be compatible with the approximate bounds on ${\Delta\mu}/{\mu}$ of at most 10 ppm. We will keep in mind these two possibilities in our analysis\,\footnote{We note that despite the fact that stable leptons (essentially electrons) do not contribute in any significant way to the r.h.s. of equations\,(\ref{mass2}) and (\ref{mass3}), the relative variation $\Delta m_e/m_e$ could be as big as $\Delta m_p/m_p$, in principle. This option must be kept in mind when considering the total time variation of quantities involving a ratio of baryon and lepton masses, such as $\mu=m_p/m_e$.}.  Both of them, however, lead to $\nu_X\sim\nu_m\sim 10^{-3}$ via Eq.\,(\ref{mass3}), what clearly points to the crucial role of the DM contribution to explain the bulk of the mass drift rate in the Universe (cf. Fig. 4).

Assuming that the anomaly indices for matter non-conservation are constant, we can integrate Eq.\,(\ref{indices}) and we find the evolution of the baryons and DM particle masses. We may most conveniently perform the integration in terms of the redshift using Eq.\,(\ref{eq:dotffp}), and we find:
\begin{equation}
\label{mass}
 m_i(z)=m_{i0}(1+z)^{-3\nu_i}  \,\,\,\,  \longrightarrow   \,\,\,\,  \frac{\Delta m_i(z)}{m_i} \simeq -3 \nu_i \ln(1+z)\,.
\end{equation}
Here we have defined $\Delta m_i(z) = m_i(z) - m_0$, with $m_0\equiv m(z=0)$; and the index $i = B, L, X$ runs over stable baryons, leptons and DM particles, respectively. The form (\ref{mass}) obtained for the mass variation of the particles with the redshift is indeed the one announced  in (\ref{eq:mz}). The total mass variation of the Universe is conceived here as a physical process connected with the variation of the particle masses themselves rather than the appearance or disappearance of new particles during the expansion process. In our framework the particle mass changes are possible thanks to the interaction with the dynamical vacuum and/or the evolution of the gravitational constant, as described in the previous section. In Fig. 4 we plot their evolution with the redshift, showing that the DM mass drift is the dominant one.

\begin{figure}
\begin{center}
\includegraphics[width=4in, height=2.5in]{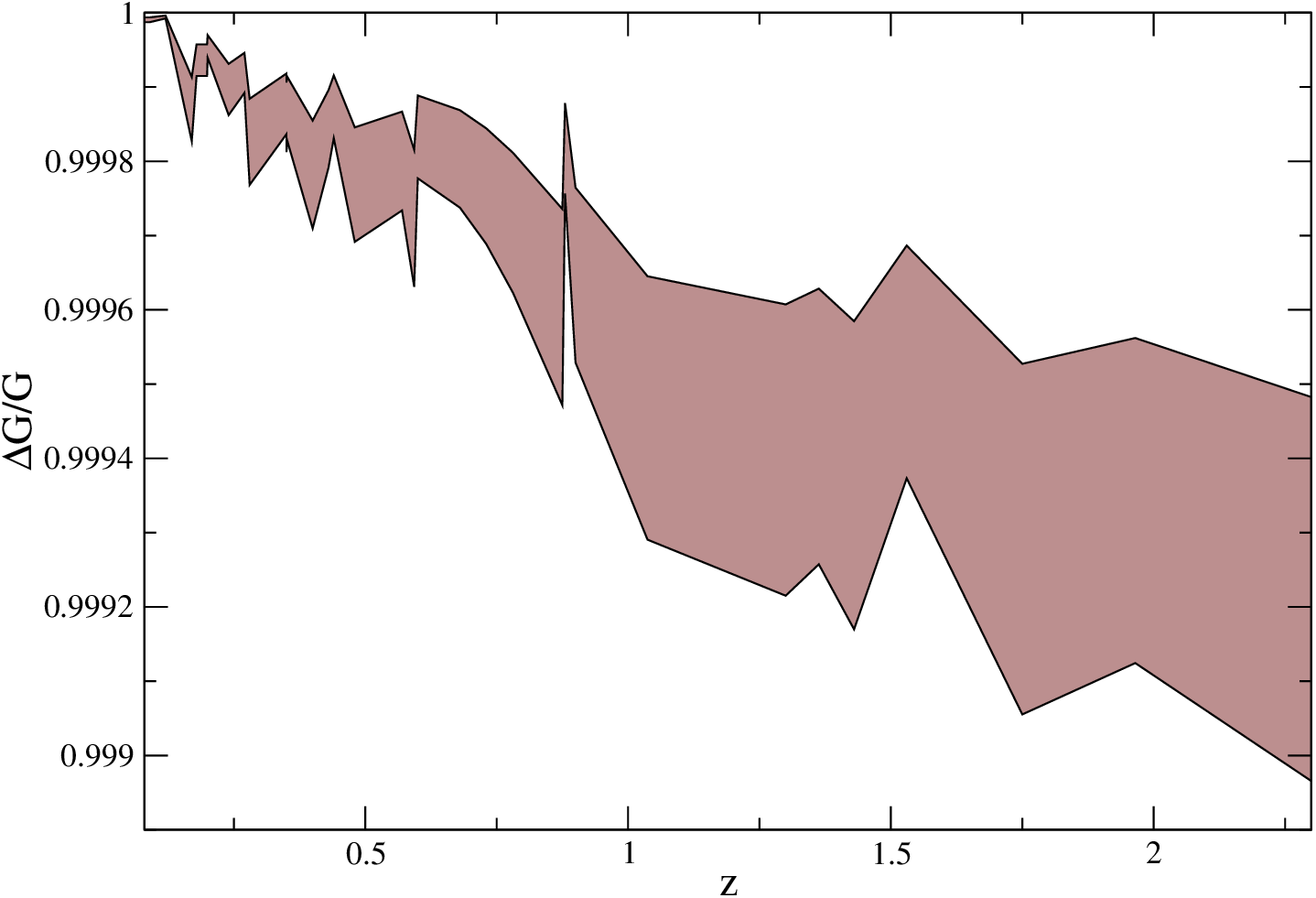}
\caption{The relative variation  $\Delta G/G$ as a function of the redshift. We display  the (shaded) region comprised in between the fit values  $\nu_G \in [0.0005;0.001]$, which encompass the typical parameter range found in the analysis of \,\cite{SolaGomCruzApJL,SolaGomCruzApJ,PRL2017}.}
\end{center}
\label{G2a}
\end{figure}

\subsection{Time variation of $G$ and $\CC$}

The corresponding cosmic drift rates of the vacuum energy density and gravitational coupling ensue from (\ref{Gevolution}) and (\ref{VacuumEvolution}). Using Eq.\,(\ref{eq:dotffp}) we find, in leading order:
\begin{equation}
\label{eq:dotrL}
\frac{\dot{\rho_{\Lambda}}}{\rho_{\Lambda}} = - 3 (\nu_m + \nu_G) \frac{\Omega_{m}}{\Omega_{\Lambda}} (1+z)^{3} H
\end{equation}
and
\begin{equation}
\label{eq:dotG}
\frac{\dot{G}}{G} = - 2 \nu_G \frac{\dot{H}}{H}=2 \nu_G \,(1+z)\,H'(z)\,,
\end{equation}
where the Hubble function for the RVM is given by Eq.\,(\ref{eq:HzRVM}).
It is convenient to trade the derivative of the Hubble function in the equation above in terms of the Hubble function itself. After some rearrangement we find:
\begin{equation}\label{eq:dotGsmaex}
\frac{\dot{G}}{G}=3\nu_G\,\Omega_{m}\,\frac{H_0}{E}\left(1+\frac{E^2-1}{\Omega_m}\right)=3\nu_G\,H\,\left(1-\frac{H_0^2}{H^2}\,\Omega_{\CC}\right)\,,
\end{equation}
which is valid for any value of the redshift.  While for small values $z<1$   the previous expression behaves roughly as $\frac{\dot{G}}{G}\simeq 3\nu_G\Omega_m\,(1+z)^3\,H_0$, for large values of the redshift we have $\frac{\dot{G}}{G}\simeq 3\nu_G\,H$.  The corresponding plots of $G(z)/G_0$ and of the cosmic drift rate $\dot{G}/G$ as a function of the redshift are depicted in Fig. 5.  We can see from the plot on the left in that figure that the value of $G$ decreases with the redshift and therefore $G$ behaves as an asymptotically free coupling, that is to say $G$ decreases in the past, which is the epoch where the Hubble rate (with natural dimension of energy) is higher. This is of course already obvious from Eq.\,(\ref{Gevolution}) for $\nu_G>0$. Moreover from the plot on the right in Fig. 5 we learn that the rhythm of variation of $G$ slows down with the cosmic expansion, i.e. the rate of change is larger in the past.
\begin{figure}
\begin{center}
\includegraphics[width=3in, height=3in]{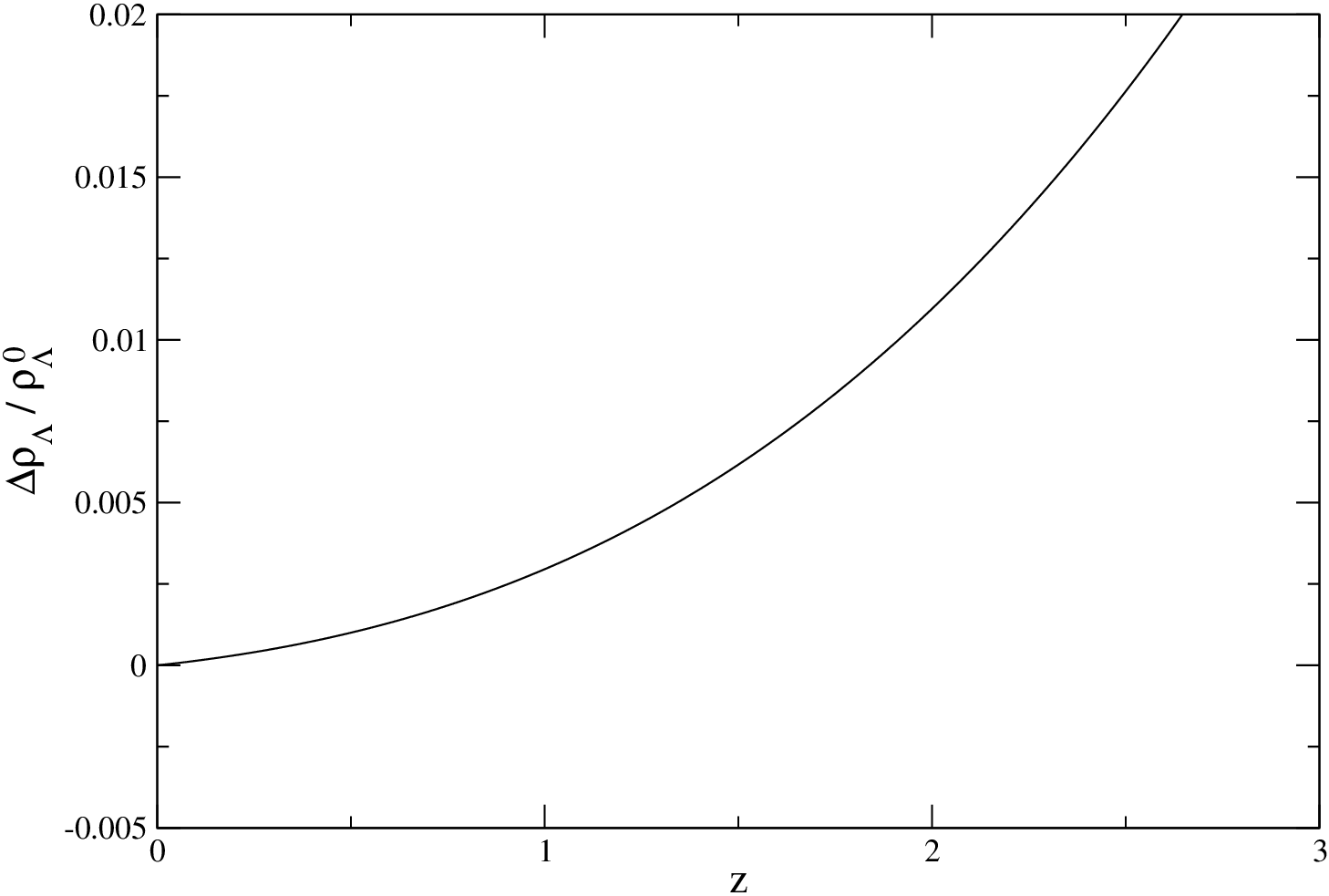}
\caption{The relative variation of $\rL(z)$ (the vacuum energy density) with respect to the current value, within the RVM for $\nu=0.001$, as a function of the redshift -- see Eq.\,(\ref{VacuumEvolution2}).}
\end{center}
\label{G2b}
\end{figure}


Following the fitting results to the overall cosmological observables obtained in \cite{SolaGomCruzApJL,SolaGomCruzApJ,GomSolaBas15} we can explore the parameter $\nu_G$ in the typical range from $0.0005$ to $0.001$ and then evaluate the relative variation  $\Delta G/G$ as a function of the redshift in that interval. The plot is depicted in Fig. 6. The larger is $\nu$ the smaller is $G$ at any given redshift. The upper curve in Fig. 6 corresponds to $\nu_G=0.0005$ whereas the lower one to $\nu_G=0.001$. The shaded area gives the prediction of $\Delta G/G$ for the values of $\nu_G$ comprised in that interval for each $z$. Worth noticing is that if we extend the domain of applicability of Eq.\,(\ref{Gevolution}) up to the BBN epoch ($z=z_N\sim 10^9$), we find that the relative variation of $G$ at the BBN time as compared to the present is $(G(z_N)-G_0)/G_0\simeq -0.06$, which is less than $10\%$ in absolute value and hence compatible with the current BBN bounds\,\cite{Uzan2011}.

Finally, in Fig. 7 we display the evolution of the vacuum energy density with the redshift according to the RVM formula (\ref{VacuumEvolution}), specifically the solid line corresponds to the relative correction with respect to the current value:
\begin{equation}
\label{VacuumEvolution2}
 \frac{\Delta\rL(z)}{\rLo}\equiv\frac{\rL(z)-\rLo}{\rLo}= \frac{\nu}{1 -\nu}\, \frac{\Omega_m}{\Omega_{\CC}}\, [(1+z)^{3(1-\nu_m)} -1]\,,
\end{equation}
where $\nu=\nu_m+\nu_G$.
We may use Eq.\,(\ref{eq:ModelAgeneral}) and rewrite the evolution of the vacuum energy density as
\begin{equation}
\label{VacuumRelativeEvolution}
 \frac{\Delta\rL(H)}{\rLo}\equiv\frac{\rL(H)-\rLo}{\rLo}= \frac{\nu}{\Omega_{\CC}}\,\left( E^2-1\right)\,.
\end{equation}
This formula displays the relative variation of the vacuum energy density with respect to the current value directly in terms of the Hubble function. In the RVM this expression is more fundamental than (\ref{VacuumEvolution2}) because it is the solution of the RGE in terms of $H$. As we shall see in Sect. \ref{sec:Alternative Models} , there are alternatives models that may lead to Eq. (\ref{VacuumEvolution2}) using ad hoc assumptions on the interaction between DE and DM. By the same token the expression of $G$ directly in terms of $H$ is closer to the spirit of the RVM since it can be derived from the RG formalism of QFT in curved spacetime\,\cite{JSPRev2013,Fossil07}.

\subsection{Time evolution of the fine-structure constant}\label{sect:alphaem}

Motivated by a possible indication of a variation (decrease) in the fine-structure constant at high redshift, as well as a possible spatial variation (see \cite{alpha0} and references therein, as well as the reviews \,\cite{Uzan2011,Calmet2015})), we will address here this topic from the point of view of the implications of the running vacuum energy density throughout the cosmic history. We have mentioned before that in the electroweak sector of the SM is not possible to establish a connection between the cosmological evolution of the weak and electromagnetic couplings to the particle masses because there is no analogue in this sector of the QCD scale parameter $\LQCD$. Notwithstanding, it is still possible to relate the electroweak couplings to $\LQCD$ itself in an indirect way if we use the hypothesis of Grand Unification of the SM couplings at a very high energy scale.  We will focus here on the fine-structure constant $\alpha_{\rm em}=e^2/4\pi$ and its correlated time-evolution with the strong coupling counterpart $\alpha_{s}=g_s^2/4\pi$, and ultimately with the time evolution of $\LQCD$ and $\mu=m_p/m_e$.
\\

In a Grand Unified Theory (GUT), the gauge couplings, and in particular the strong gauge coupling given in Eq. (\ref{eq:alphasLQCD}), can be made to converge to a unification point (at a high energy scale $M_X$) with the electroweak couplings. This is possible if more matter content is appropriately added (e.g. from supersymmetric particles), in which case $M_X\sim 10^{16}$ GeV\,\cite{GUT}. This feature can be used as a theoretical argument to connect the possible time variation of the running coupling constants\,\cite{FritzschSola2012,CalmetFritzsch}.
Let $d\alpha_i/dz$ be the variation of $\alpha_i$ with the cosmological redshift $z$. Such variation is possible if we have a consistent theoretical framework supporting this possibility, such as the RVM picture described in sect.\ref{sec:RVM}. Each of the couplings $\alpha_i=g^2_i/4\pi$
is a function of the running scale $\mu_R$, and they follow the standard (1-loop) running laws
\begin{equation}\label{eq:GUTrunning}
\frac{1}{\alpha_i(\mu_R,z)}=\frac{1}{\alpha_i(\mu'_R,z)}+\frac{b_i}{2\pi}\ln\frac{\mu'_R}{\mu_R}\,,
\end{equation}
to which we have appended the redshift variable to parameterize the cosmic evolution. Since the $\beta$-function coefficients $b_i$ of the running are constant in time and redshift, it follows that the expression $\alpha_i'(z)/\alpha_i^2(z)\equiv(d\alpha_i /dz)/\alpha^2_i$ is independent of $\mu_R$, i.e. it is a RG-invariant.
Using this property and our ansatz concerning the cosmological evolution of the particle masses in the RVM, one can show that the running of the electromagnetic coupling $\alpha_{em}$ is related to the corresponding cosmic running of the strong coupling $\alpha_{s}$  as follows\,\,\cite{FritzschSola2012,CalmetFritzsch}:
\begin{equation}
\frac{1}{\alpha_{em}} \frac{d \alpha_{em} (\mu_R,z)}{dz} = \frac{8}{3} \frac{\alpha_{em}(\mu_R,z)}{\alpha_{s}(\mu_R,z)}\frac{1}{\alpha_{s}} \frac{d \alpha_{s} (\mu_R,z)}{dz}\,.
\end{equation}
Combining this expression with Eq.\,(\ref{alpha_s}) we can reexpress the cosmic running of ${\alpha_{em}}$ in terms of the cosmic running of the QCD scale, and  we find:
\begin{equation}
\frac{1}{\alpha_{em}} \frac{d \alpha_{em} (\mu_R,z)}{dz} = \frac{8}{3} \frac{\alpha_{em}(\mu_R,z)/\alpha_{s}(\mu_R,z)}{\ln(\mu_R/\Lambda_{QCD})}\frac{1}{\LQCD}\frac{d \LQCD(z)}{dz}.
\end{equation}
At the $Z$-boson mass scale $\mu_R = M_Z$, where both $\alpha_{em}$ and $\alpha_s$ are known with precision, one obtains
\begin{equation}\label{eq:alphaemLQCD}
\frac{1}{\alpha_{em}} \frac{d \alpha_{em} (\mu_R,z)}{dz} \simeq 0.03 \frac{1}{\LQCD}\frac{d \LQCD(z)}{dz}.
\end{equation}
The above equation can now be nicely connected with our discussion of the cosmic running of the particle masses considered in sect.\,\ref{sec:RunMass}. Indeed, we have seen that the proton mass receives the bulk of its contribution from $\LQCD$ through $m_p\simeq c_{QCD}\,\LQCD$ (with a negligible contribution from the quark masses and the electromagnetism). Therefore, substituting this expression in Eq.\,(\ref{eq:alphaemLQCD}) and integrating,  and then inserting the redshift dependence of the proton mass through (\ref{mass}), we find:
\begin{equation}\label{eq>alphaemz}
\alpha_{\rm em}(z)\simeq \alpha_{\rm em}^0\left(\frac{m_p(z)}{m_p^0}\right)^{0.03}=\alpha_{\rm em}^0\,(1+z)^{-0.09\nu_B}\,,
\end{equation}
where $\alpha_{\rm em}(z)$ stands for the value of the fine structure constant at redshift $z$ at a fixed value of $\mu_R$, and $\alpha_{\rm em}^0(z)$ is its current value ($z=0$). Since $\nu_B$ is a small parameter, related to the fitted value $\nu_m\sim 10^{-3}$\,\cite{SolaGomCruzApJL,SolaGomCruzApJ,PRL2017} through (\ref{mass3}), we can estimate the relative variation of the electromagnetic fine structure constant with the redshift as follows:
\begin{equation}\label{eq:alphavariation}
\frac{\Delta\alpha_{\rm em}(z)}{\alpha_{\rm em}}\simeq -0.09\,\nu_B\,\ln(1+z)\,.
\end{equation}
\begin{table}[!t]
      \begin{center}
          \begin{tabular}{cccc}
          \hline
          \hline
               &$ z $&$ \Delta \alpha/ \alpha (ppm) $& Ref.\\
          \hline
          \hline
               &$ 1.08 $&$ 4.3 \pm 3.4$& \cite{alpha1}\\
               &$ 1.14 $&$ -7.5 \pm 5.5$& \cite{alpha2}\\
               &$ 1.15 $&$ -0.1 \pm 1.8$& \cite{alpha3}\\
               &$ 1.15 $&$ 0.5 \pm 2.4$& \cite{alpha4}\\
               &$ 1.34 $&$ -0.7 \pm 6.6$& \cite{alpha2}\\
               &$ 1.58 $&$ -1.5 \pm 2.6$& \cite{alpha6}\\
               &$ 1.66 $&$ -4.7 \pm 5.3$& \cite{alpha1}\\
               &$ 1.69 $&$ 1.3 \pm 2.6$& \cite{alpha5}\\
               &$ 1.74 $&$ -7.9 \pm 6.2$& \cite{alpha1}\\
               &$ 1.80 $&$ -6.4 \pm 7.2$& \cite{alpha2}\\
               &$ 1.84 $&$ 5.7 \pm 2.7$& \cite{alpha3}\\
          \hline
          \hline
          \end{tabular}
      \end{center}
      \caption{Compilation of recent direct measurements of the fine-structure constant obtained
                by different spectrographic methods. For details of these methods, see the references cited above. }
      \label{tab1}
\end{table}
Defining $\Delta\alpha_{\rm em}/\alpha_{\rm em}=3\nu_{\rm em}$ by analogy  with Eq.\,(\ref{eq:variationmi}), we learn that the effective running index of the em coupling is some $30$ times smaller than that of the baryonic index and with opposite sign, in other words $\nu_{\rm em}\simeq -0.03\,\nu_B$ (up to logarithmic evolution with the redshift).

Let us now consider the observational situation concerning the measurements of the possible time variation of $\alpha_{\rm em}$ and the implied restrictions on the parameter $\nu_B$ from Eq.\,(\ref{eq:alphavariation}).
Table 1 shows recent measurements of $\alpha_{em}$ in the redshift range $ 1.08 \leq z \leq 1.84$. In Fig.\,8 we have plotted these measurements. Let us also mention that apart from the astrophysical measurements of the time variation of $\alem$, which give access to large look-back times of order of few billion years from now (corresponding to large redshifts of the order of those indicated in Table 1), there is a parallel research line of high precision laboratory based measurements whose look-back times are of course necessarily much more modest, but whose outstanding precision (based on state-of-art quantum optic techniques involving atomic clocks) can be highly competitive\,\cite{FritzschBook}. For instance, by comparing hyperfine transitions of different chemical elements with the cesium atomic clock, one can derive constrains on the time variation of the fundamental constants. Usually these measurements are correlated with the time dependence of the ratio $\mu=m_p/m_e$ through the the nuclear magnetic moment. In all these cases the typical result within errors (mostly compatible with zero) is $|\Delta\dot{\alpha}_{\rm em}/\alem|\lesssim 10^{-17}-10^{-16}$\,yr$^{-1}$. These correspond once more to  an upper bound on a relative variation  $|\Delta{\alpha}_{\rm em}/\alem|$ of order of $1-10$ ppm, see \cite{Uzan2011,Calmet2015}, and therefore competitive with the astrophysical measurements. A particular lab experiment\,\cite{Rosenband2008} employing narrow optical transitions in $Hg^+$ and $Al^+$
ions was able to directly measure changes in $\alem$ independent of other parameters. Comparing the transition frequencies over 12 months, the experiment renders a drift rate of $\Delta\dot{\alpha}_{\rm em}/\alem= (-1.6\pm 2.3) 10^{-17}$\,yr$^{-1}$, thus providing an upper bound on $|\Delta\alpha_{\rm em}/\alpha_{\rm em}|$ of order of a few ppm.

\begin{figure}
\begin{center}
\label{fig:alphavalues}
\includegraphics[width=4in, height=3in]{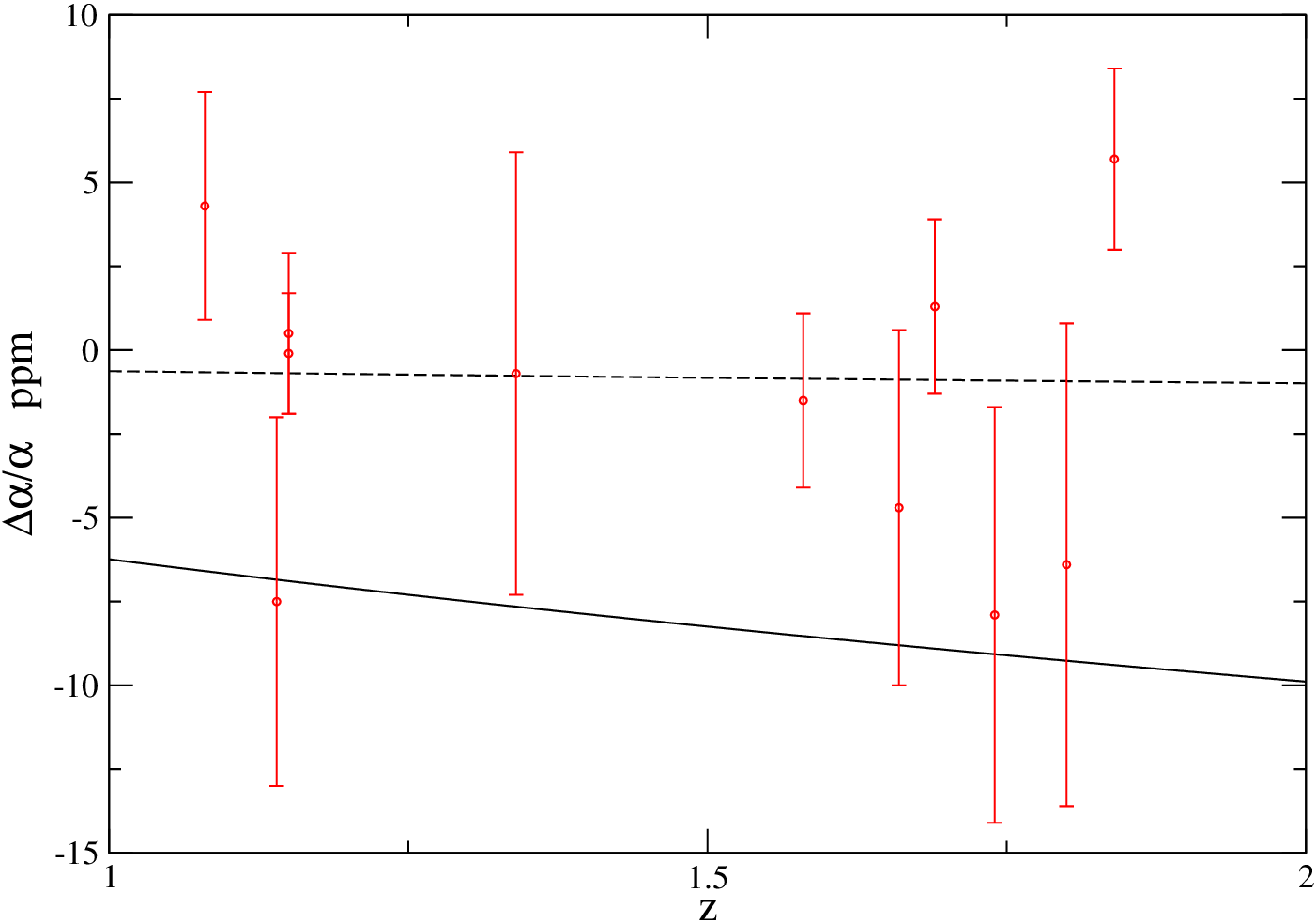}
\caption{The data points of Table 1 on the relative variation $\Delta\alem/\alem$ at different redshifts (in ppm).
The solid and dashed lines correspond to the theoretical combined RVM-GUT prediction based on formula
(\ref{eq:alphavariation}) for the values $\nu_B=10^{-4}$ and $10^{-5}$, respectively.
The tendency of the data to reflect smaller values of $\alem$ at large $z$ is correctly described by the theoretical curves (which indicate $\Delta\alpha_{\rm em}<0$),
although the current observational errors are still too large.}
\end{center}
\end{figure}

What is the possible impact of the RVM here? It is remarkable that the above mentioned results, whether from astrophysical or lab measurements, can be accounted for (in order of magnitude) within the RVM in combination with the GUT hypothesis. Indeed, we can see that the  theoretical RVM prediction falls right within the order of magnitude of the typical measurements quoted in Table 1  and in Fig.\,8, provided $\nu_B$ lies in the range from $10^{-4}$ to $10^{-5}$.  This follows from Eq.(\ref{eq:alphavariation}), which, roughly speaking, says that the RVM prediction is of order $\Delta\alpha_{\rm em}/\alpha_{\rm em}\sim -0.09\nu_B$ up to log corrections in the redshift. More precisely, in Fig.\,8 we have superimposed the exact theoretical prediction $\alpha_{\rm em}(z)$  according to the formula (\ref{eq:alphavariation}). We can see that, notwithstanding the sizeable error bars, the trend of the measurements in Table 1 suggests a decrease of $\alpha_{\rm em}$ with the redshift (as there are more points compatible with $\Delta\alem<0$ than points compatible with $\Delta\alem>0$). This behavior has been previously noted in the literature\,\cite{alpha0} and is roughly in accordance with our theoretical curves  in Fig.\,8. But of course we need more precise measurements to confirm the real tendency of the data, as the errors are still too large and no firm conclusion is currently possible.

The following remarks should be emphasized. First, despite it is not possible to be more precise concerning the best fit value for $\nu_B$, in all cases the measurements in Table 1 indicate a maximum effect of $1-10$  ppm, i.e. $|\Delta\alpha_{\rm em}/\alpha_{\rm em}|$ at the level of $10^{-6}$ to $10^{-5}$.  This can be accommodated in the RVM framework since $\nu_B$ describes the effect from only the baryonic component in Eq.\,(\ref{mass3}). Such component should be naturally smaller than the value of the total index $\nu_m\sim 10^{-3}$ fitted on the basis of the overall analysis involving the SNIa+BAO+H(z)+LSS+BBN+CMB observables\,\cite{SolaGomCruzApJL,SolaGomCruzApJ,GomSolaBas15}.
Second, the correct order of magnitude for $\nu_B$, which we have obtained from the direct $\Delta\alem/\alem$ observations (viz. $\nu_B\sim 10^{-4}-10^{-5}$) does coincide with the result inferred from our previous considerations on the alternative observable $\Delta\mu/\mu$ in sect.\,\ref{sec:RunMass}. Put another way, we could have input the value of $\nu_B$ needed to explain the typical measurements of the observable $\Delta\mu/\mu$ and we would have naturally predicted the typical range of values of $\Delta\alem/\alem$ derived from direct observations, and vice versa. As noted previously, this is because the RVM in combination with the GUT framework neatly predicts the relation $\nu_{\rm em}\simeq -0.03\,\nu_B$.

In the light of the above results, the baryonic index $\nu_B$ in Eq.\,(\ref{mass3}) is definitely subdominant as compared to the dark matter one, $|\nu_B|\ll|\nu_X|$, and hence $\nu_X$ must be of order of the total matter index $\nu_m\sim 10^{-3}$ fitted from the overall cosmological observations\,\cite{SolaGomCruzApJL,SolaGomCruzApJ,GomSolaBas15}. In other words, we find once more that it must be the DM component that provides the bulk of the contribution to the time variation of masses in the Universe. This fact was not obvious a priori, and is not necessarily related to the overwhelming abundance of DM as compared to baryons, for the large amounts of DM could simply remain passive and not evolve at all throughout the cosmic expansion.  If the best fit value of the total mass variation index $\nu_m$ would have been, say of order $10^{-5}$, equation (\ref{mass3}) could have been naturally fulfilled with $\nu_X<<\nu_B\sim 10^{-4}$ and this would still be compatible with the measurements in Table 1.  However, the fact that the value of $\nu_m$ (obtained from the overall cosmological fit to the data within the RVM\,\cite{SolaGomCruzApJL,SolaGomCruzApJ,PRL2017}) comes out significantly larger than the baryonic index $\nu_B$ has nontrivial consequences and provides an independent hint of the need for  (time-evolving) dark matter. Taking into account that we have been able to infer this same conclusion from the analysis of the two independent observables $\Delta\mu/\mu$ and $\Delta\alem/\alem$, which become correlated in this theoretical framework, does reinforce the RVM scenario and places the contribution from the DM component to the forefront of our considerations concerning the total mass drift rate in the Universe\,\cite{FritzschSola2012}.

\section{Alternative dynamical vacuum models interacting with matter}
\label{sec:Alternative Models}

Let us finally address some alternative scenarios for the non-conservation of the particle masses in an expanding universe.
Our starting point is a phenomenological coupling between the vacuum energy density and the matter density, where for simplicity we now assume $G=$const. The background evolution is then encoded in the coupled system of local energy conservation equations
\begin{equation}
\label{darkQ1}
 \dot{\rho}_m + 3H\rho_m = Q
\end{equation}
and
\begin{equation}\label{darkQ2}
 \dot{\rho_{\Lambda}} = - Q,
\end{equation}
where $Q$ denotes the background energy source between dark matter and vacuum energy (or in general some dark energy source). From the previous equations we see that for $Q > 0$ the matter energy density increases whereas the vacuum energy density decreases, and hence the energy flows from vacuum to matter, and vice versa for $Q$ featuring the opposite sign. In other words, for $Q>0$ the vacuum is decaying into matter whereas for $Q<0$ the matter decays into vacuum. Of these two options the naturally preferred one, at least from the point of view of the second principle of thermodynamics, should be the first one. Despite it is usually assumed that the vacuum decays only into DM, this effect has little quantitative implications for the present analysis since we have seen that the baryonic component is essentially conserved ($\xi_B<<\xi_X$, see sect. 4.1). Therefore we shall not consider this correction here. For more details, see \cite{JSPReview2016}.

Many functions  can be proposed for the interacting energy source $Q$, see e.g. \cite{Bolotin2015,Salvatelli2014,Murgia2016}. For our purposes, it will suffice to consider two frequently discussed phenomenological expressions in the literature, namely a source proportional to $\rL$ in the form $Q = \qL H \rho_{\Lambda}$ (hereafter called ``$\qL$-model''), and a source proportional to the matter density, $Q = q_m H \rho_{m}$ (``$q_m$-model''), where $\qL$ and $q_m$ are small dimensionless parameters ($|\qL|,|q_m|\ll1$). Considering the $\qL$-model, we easily find from (\ref{darkQ2}) the explicit form of the vacuum evolution law
\begin{equation}\label{eq:rLaSalva}
\rL(a)=\rLo\,a^{-\qL}\,,
\end{equation}
which can be substituted in (\ref{darkQ1}) and upon integration we derive the corresponding matter density evolution in that model:
\begin{equation}\label{eq:rmSalva}
\rho_m(a)=\rho_m^0\,a^{-3}-\frac{\qL}{3-\qL}\,\rLo\left(a^{-3}-a^{-\qL}\right)\,.
\end{equation}
Obviously equations (\ref{eq:rLaSalva}) and (\ref{eq:rmSalva}) are very different from the corresponding ones from the RVM, see e.g. equations (\ref{MatterEvolution}) and (\ref{VacuumEvolution}) for $\nu_G=0$ (since we are now considering $G=$const.). With the help of these results and assuming that the number density of particles is conserved, i.e., $n(a)=n_0\,a^{-3}$ (cf. sect. 3), we can obtain the law for the cosmological evolution of masses. Notice that $\rho_{m}(a)=n(a) m(a)$ and $\rho_{m}^0=n_0\,m_0$.
Phrased in terms of the redshift, the  relative mass variation can be cast as follows:
\begin{equation}
\label{dark_matter4}
\frac{\Delta m(z)}{m_0}\equiv \frac{m(z)-m_0}{m_0}= \frac{\qL}{3-\qL}\,\frac{\Omega_{\CC}}{\Omega_m}\left[(1+z)^{\qL-3}-1\right]\,.
\end{equation}
Notice that $\delta m/\delta z=-\qL\,m_0\left(\Omega_{\CC}/\Omega_m\right)(1+z)^{\qL-4}$, so that the mass tends to decrease or increase with the redshift (equivalently, increase or decrease with the expansion) for $\qL>0$ or $\qL<0$ respectively (recall that $|\qL|\ll1$). In the remote past ($z>>1$), we have the initial value $m_i\simeq m_0\left(1-\qL\Omega_{\CC}/3\Omega_m\right)$, whereas in the remote future ($z\to-1$, i.e. $a\to\infty$) the asymptotic final mass value evolve as $m_f\simeq m_0 \left(\qL\Omega_{\CC}/3\Omega_m\right)a^{3-\qL}$. For $\qL<0$ we have $Q<0$, which we already pointed out as being unfavored by the second law of thermodynamics since the energy flows from matter into vacuum. Even worse, for $\qL<0$ the asymptotic mass values $m_f$ become eventually negative, which would correspond to a rather unstable situation for the Universe. The lack of sympathy for this sign is explicitly confirmed by the analysis of the current observational data when confronted with the various types of dynamical vacuum models. One finds that $\qL<0$ gives a very bad fit as compared to $\qL>0$, see the study of Ref.\,\cite{PRL2017}.

Let us next assess the situation with the second alternative model for the dark energy source mentioned above, $Q = q_m H \rho_m$ (the $q_m$-model). We can easily integrate equations (\ref{darkQ1}) and (\ref{darkQ2}) anew, with the following results:
\begin{equation}\label{eq:Massqm}
\rho_m(a)=\rho_m^0\,a^{-3+q_m}
\end{equation}
and
\begin{equation}
\label{eq:Vacuumqm}
 \rL(a)=\rLo+\frac{q_m\,\rho_m^0}{3-q_m}\left(a^{-3+q_m}-1\right)\,.
\end{equation}
Clearly these expressions are formally similar to the corresponding ones  in the RVM, i.e. equations (\ref{MatterEvolution}) and (\ref{VacuumEvolution}), with the identification $q_m=3\nu_m$ and taking into account that $\nu_G=0$ here (i.e. $G$ is constant). The mass evolution with the redshift is therefore as in Eq.\,(\ref{eq:mz}), i.e.
\begin{equation}\label{eq:qmMass}
m(z)=m_0\,(1+z)^{-q_m}\,.
\end{equation}
We find that $m(z)$ decreases with $z$, and hence increases with the expansion, if $q_m>0$ (corresponding to a situation of decay of vacuum into matter); and $m(z)$ decreases with the expansion if $q_m<0$ (when matter decays into vacuum).  In contradistinction to the $q_{\CC}$-model, we note that the masses remain now always positive irrespective of the sign of $q_m$.

We point out that in the two models considered in this section (the $\qL$-model and the $q_m$-model) there exists no special motivation for the particular form they have for the interacting source $Q$. In contrast, in the RVM case the form of $Q$ is not taken as a mere phenomenological ansatz, it  is theoretically determined. In fact, it can be derived from the dynamical vacuum equation (\ref{eq:ModelAgeneral}), which leads to Eq.\, (\ref{VacuumEvolution}) when the model is explicitly solved. For example, take  $G=$const. ($\nu_G=0$) in which case the local conservation equation for the RVM can be put in the form (\ref{darkQ1}) with $Q=-\dot{\rho}_{\Lambda}$. We may compute explicitly the time derivative of $\rL$ and reexpress the result in terms of the redshift using Eq.\,(\ref{eq:dotffp}). We find:
\begin{equation}\label{eq:QRVM}
Q=-\dot{\rho}_{\Lambda}=\rho_{\CC}'(z)\,(1+z)\,H=3\,\nu_m\rho_m\,H\,,
\end{equation}
where use has been made of the expressions for  $\rho_m(z)$ and $\rL(z)$ given in equations\ (\ref{MatterEvolution}) and(\ref{VacuumEvolution}) with $\nu_G=0$. Alternatively, we may use the generalized conservation law Eq.\,(\ref{GeneralizedConservation}) with $G'(z)=0$; and we find, once more,  $Q=\rho_{\CC}'(z)\,(1+z)\,H=(3\rho_m(z)-\rho'_m(z)(1+z))H=+3\,\nu_m\rho_m\,H$. In either way we arrive at
\begin{equation}\label{darkQ1L}
 \dot{\rho}_m + 3H\rho_m = -\dot{\rho}_{\Lambda}=+3\,\nu_m\rho_m\,H\,.
\end{equation}
This equation clearly shows that the RVM with $G=$const. behaves effectively (at least in the matter-dominated epoch) as a $q_m$-model with $q_m=3\nu_m$. When radiation is included the two models present significant differences. These differences have been accounted for in \cite{PRL2017,JSPReview2016} and they have implications for the overall fit to the data, especially for the high redshift data of course. The net outcome is that the RVM does better than the $q_m$-model, but both of them do better than the $\CC$CDM, see\,\cite{JSPReview2016} for details.

Recall that ultimately the behavior of the RVM stems from the dynamical vacuum equation (\ref{eq:ModelAgeneral}), which follows from the RG-flow in QFT in curved spacetime\,\cite{JSPRev2013,SolaGomezReview}. As mentioned in sect.\,3,  Eq.\,(\ref{eq:ModelAgeneral}) can be extended with higher powers of $H$ so as to include inflation in a single unified theory describing the cosmic evolution from the early universe until the current one\,\cite{SolaGomezReview,GRF2015,LBS1,LBS2,LBS3}.

From the foregoing discussion it is clear that different dynamical vacuum energy models exist for describing the possible time variation of the fundamental constants in a framework which is consistent with GR. Some of these models are more phenomenological, but in the RVM case there is a more concrete theoretical motivation for the dynamical vacuum structure. It suggests that a slow change in the
fundamental ``constants'' can be theoretically motivated and cannot be ruled out at present. The subject is therefore worthwhile being further investigated in the light of new data, as it can reveal new clues to fundamental physics.

\section{Discussion and conclusions}
\label{sec:Conclusions}

In this paper we have addressed the running vacuum models (RVM) of the cosmic evolution and the possible implications they could have in explaining the reported hints of the time variation of the so-called fundamental constants of Nature, such as masses, coupling constants etc, including the gravitational coupling $G$ and the $\CC$-term in Einstein's equations.  The impact from the RVM on this issue stems from the cosmological energy exchange between vacuum, matter and the possible interplay with the Newtonian coupling $G$ and the vacuum energy density $\rL=\CC/8\pi G$. Because the possible cosmological running of these quantities is controlled by the Hubble parameter, the RVM predicts that the associated rhythm of change, i.e. the drift rate of the fundamental constants should naturally be as moderate as dictated by the expansion rate of the Universe at any given instant of the cosmic history. On this basis it is possible to compute the time evolution of the vacuum energy density and the corresponding change of the gravitational coupling and the particle masses. Combining this scenario with the GUT hypothesis we have obtained a prediction for the time evolution of the fine structure constant correlated with the time evolution of the proton mass, or more precisely the proton-to-electron mass ratio $\mu=m_p/m_e$.

Taking into account that the small, but nonvanishing, rate of change of the vacuum energy density has been fitted to the cosmological data at a rather significant confidence level -- see the recent studies \,\cite{SolaGomCruzApJL,SolaGomCruzApJ,PRL2017,SolaGomCruzMPLA} -- and baring in mind that such vacuum rate of change impinges on the corresponding variation of the particle masses, we conclude that the mass variation must be essentially supported by the mass drift rate of the dark matter (DM) particles (since the time evolution of the baryonic component is found to be some two orders of magnitude smaller). This can be interpreted as an indirect alternative hint of the need for DM.
If in the future the precision of these experiments further improves we might well find ourselves on the verge of measuring these subtle effects and perhaps be in position to check if they can be explained within the kind of theoretical running vacuum models that we have studied here.

\vspace{0.5cm}


\noindent{\bf Acknowledgments:} \vspace{0.2cm}

HF and JS are both grateful to the Institute for Advanced Study at the Nanyang Technological University in Singapore for hospitality
and support while this work was being accomplished. JS has been supported in part by FPA2013-46570 (MICINN), CSD2007-00042 (CPAN), 2014-SGR-104 (Generalitat de Catalunya) and MDM-2014-0369 (ICCUB). RCN acknowledges financial support from CAPES Foundation Grant No. 13222/13-9 and is grateful for the hospitality at the Dept. FQA, Universitat de Barcelona.

\newcommand{\JHEP}[3]{ {JHEP} {#1} (#2)  {#3}}
\newcommand{\NPB}[3]{{ Nucl. Phys. } {\bf B#1} (#2)  {#3}}
\newcommand{\NPPS}[3]{{ Nucl. Phys. Proc. Supp. } {\bf #1} (#2)  {#3}}
\newcommand{\PRD}[3]{{ Phys. Rev. } {\bf D#1} (#2)   {#3}}
\newcommand{\PLB}[3]{{ Phys. Lett. } {\bf B#1} (#2)  {#3}}
\newcommand{\EPJ}[3]{{ Eur. Phys. J } {\bf C#1} (#2)  {#3}}
\newcommand{\PR}[3]{{ Phys. Rep. } {\bf #1} (#2)  {#3}}
\newcommand{\RMP}[3]{{ Rev. Mod. Phys. } {\bf #1} (#2)  {#3}}
\newcommand{\IJMP}[3]{{ Int. J. of Mod. Phys. } {\bf #1} (#2)  {#3}}
\newcommand{\PRL}[3]{{ Phys. Rev. Lett. } {\bf #1} (#2) {#3}}
\newcommand{\ZFP}[3]{{ Zeitsch. f. Physik } {\bf C#1} (#2)  {#3}}
\newcommand{\MPLA}[3]{{ Mod. Phys. Lett. } {\bf A#1} (#2) {#3}}
\newcommand{\CQG}[3]{{ Class. Quant. Grav. } {\bf #1} (#2) {#3}}
\newcommand{\JCAP}[3]{{ JCAP} {\bf#1} (#2)  {#3}}
\newcommand{\APJ}[3]{{ Astrophys. J. } {\bf #1} (#2)  {#3}}
\newcommand{\AMJ}[3]{{ Astronom. J. } {\bf #1} (#2)  {#3}}
\newcommand{\APP}[3]{{ Astropart. Phys. } {\bf #1} (#2)  {#3}}
\newcommand{\AAP}[3]{{ Astron. Astrophys. } {\bf #1} (#2)  {#3}}
\newcommand{\MNRAS}[3]{{ Mon. Not. Roy. Astron. Soc.} {\bf #1} (#2)  {#3}}
\newcommand{\JPA}[3]{{ J. Phys. A: Math. Theor.} {\bf #1} (#2)  {#3}}
\newcommand{\ProgS}[3]{{ Prog. Theor. Phys. Supp.} {\bf #1} (#2)  {#3}}
\newcommand{\APJS}[3]{{ Astrophys. J. Supl.} {\bf #1} (#2)  {#3}}

\newcommand{\Prog}[3]{{ Prog. Theor. Phys.} {\bf #1}  (#2) {#3}}
\newcommand{\IJMPA}[3]{{ Int. J. of Mod. Phys. A} {\bf #1}  {(#2)} {#3}}
\newcommand{\IJMPD}[3]{{ Int. J. of Mod. Phys. D} {\bf #1}  {(#2)} {#3}}
\newcommand{\GRG}[3]{{ Gen. Rel. Grav.} {\bf #1}  {(#2)} {#3}}



{\small


\begin{thebibliography}{}


\bibitem{FritzschBook} H. Fritzsch, \textit{The Fundamental Constants:A Mystery of Physics} (World Scientific, 2009).

\bibitem{Milne1935}  E. A. Milne, \textit{Relativity, Gravitation and World Structure} (Clarendon press, Oxford, 1935); Proc. Roy. Soc. A\textbf{3} (1937) 242.

\bibitem{Dirac1937} P. A. M. Dirac,
Nature \textbf{139} (1937) 323; Proc. Roy. Soc. London A\textbf{165} (1938) 198.


\bibitem{Jordan1937} P. Jordan, Naturwiss. \textbf{25} (1937) {\bf 513}; Z. Physik {\bf 113} (1939) 660.

\bibitem{JordanBook} P. Jordan, Nature \textbf{164} (1949) 637; P. Jordan, \textit{Schwerkraft und Weltall. Grundlagen der theoretischen Kosmologie} (Friedr. Vieweg \& Sohn, Braunschweig, 1955).

\bibitem{Fierz1956} M. Fierz, Helv. Phys. Acta {\bf 29}  (1956) 128.

\bibitem{BD}  C. Brans and R. H. Dicke, Phys. Rev. {\bf 124} (1961) 925;
    R. H. Dicke, Phys. Rev. {\bf 125} (1962) 2163.

\bibitem{DickeReviews57and61} R.H. Dicke,
Rev. Mod. Phys. {\bf 29}  (1957) 355;
Nature {\bf 192} (1961) 440.

\bibitem{Damour2012} T. Damour,
Class.Quant.Grav. {\bf 29} (2012) 184001.

\bibitem{Wagoner1970}  	
P.G. Bergmann, Int. J. Theor. Phys. 1 (1968) 25;
K. Nordtvedt, Astrophys. J. 161 (1970) 1059;
R. V. Wagoner, Phys. Rev. D{\bf 1} (1970) 3209.


\bibitem{Gamow}  G. Gamow, Phys. Rev. Lett. \textbf{19} (1967) 759.

\bibitem{alphat1} J. K. Webb et al, Phys. Rev. Lett. 82 (1999) 884; J. K. Webb et al., Phys. Rev. Lett. 87 (2001) 091301; M.T. Murphy et al,
Mon.Not.Roy.Astron.Soc. 327 (2001) 1208; M. T. Murphy, J. K. Webb, V. V. Flambaum, Mon.Not.Roy.Astron.Soc. 345 (2003) 609.

\bibitem{alphat2} H. Chand, R. Srianand, P. Petitjean and B. Aracil, A\&A 417 (2004) 853; R. Srianand, H. Chand, P. Petitjean, and B. Aracil, Phys.
Rev. Lett. 92 (2004) 121302.

\bibitem{Fritzsch2009}  	
H. Fritzsch, Phys. Usp. {\bf 52} (2009) 359.

\bibitem{Barrow2010}  J. D. Barrow, Annalen Phys. \textbf{19} (2010) 202.

\bibitem{Uzan2011}  J-P. Uzan, Liv. Rev. Rel. \textbf{14} (2011) 2;
T. Chiba, Prog. Theor. Phys. \textbf{126} (2011) 993.

\bibitem{Calmet2015}  X. Calmet, M. Keller,
Mod. Phys. Lett. A{\bf 30} (2015) 1540028.

    \bibitem{Reinhold2006} E. Reinhold, R. Buning, U. Hollenstein, A. Ivanchik, P. Petitjean, W. Ubachs, Phys. Rev.
Lett. 96 (2006) 151101.


\bibitem{King2008} J. A. King, J. K. Webb, M. T. Murphy, and
R. F. Carswell, Phys. Rev. Lett. 101 (2008) 251304.


\bibitem{Preface} J. Sol\`a, Mod. Phys. Lett. A30 (2015) 1502004.

\bibitem{TimeConstants} H. Terazawa,
Phys. Lett. B{\bf 101} (1981) 43:
J. D. Bekenstein,
Phys. Rev. D{\bf 25} (1982) 1527;
W. J. Marciano,
Phys. Rev. Lett. {\bf 52}  (1984) 489; R. D. Peccei, J. Sol\`a, C.Wetterich,
Phys. Lett. B{\bf 195} (1987) 183;
T. Damour, A. M. Polyakov,
Nucl. Phys. B{\bf 423}  (1994) 532;
K. A. Olive, M. Pospelov,
Phys. Rev. D{\bf 65} (2002) 085044.

\bibitem{Flambaum2015} Y.V. Stadnik, V.V. Flambaum, Phys. Rev. Lett. {\bf 115} (2015) 201301.


\bibitem{SpecialIssueMPLA} \textit{Fundamental Constants in Physics and Their Time Variation}, Mod. Phys. Lett. A. {\bf 30} (2015), Special Issue, ed. J. Sol\`a.

\bibitem{FritzschSola2012} H. Fritzsch, J. Sol\`a,  Class. Quant. Grav. {\bf 29} (2012) 215002.

\bibitem{planck2015} P. A. R. Ade et al., (Planck Collaboration), Planck 2015 results. XIII. Cosmological parameters, Astron.Astrophys. {\bf 594} (2016) A13 [e-Print: arXiv:1502.01589].


\bibitem{MG14}  J. Sol\`a, e-Print: arXiv:1601.01668 (to appear in the proc. of the 14th Marcel Grossmann Meeting, World Scientific 2017).


\bibitem{JSPRev2013} J. Sol\`a, J. Phys. Conf. Ser. {\bf 453} (2013) 012015 [e-Print: arXiv:1306.1527]; AIP Conf.Proc. {\bf 1606} (2014) 19 [e-Print: arXiv:1402.7049].


\bibitem{SolaGomezReview} J. Sol\`a, A. G´omez-Valent, Int. J. Mod. Phys. D24 (2015) 1541003. 

\bibitem{JSPReview2016}   J. Sol\`a, Int. J. Mod. Phys. A31 (2016) 1630035.


\bibitem{GomSolaBas15} A. G\'omez-Valent, J. Sol\`a, S. Basilakos, JCAP {\bf 1501} (2015) 004.


\bibitem{GomSola2015} A. G\'omez-Valent, J. Sol\`a, Mon. Not. Roy. Astron. Soc. {\bf 448} (2015) 2810.

\bibitem{Elahe2015} A. G\'omez-Valent, E. Karimkhani, J. Sol\`a,  JCAP {\bf 1512} (2015) 048.

\bibitem{GRF2015} J. Sol\`a, Int. J. Mod. Phys. D{\bf 24} (2015) 1544027. 

\bibitem{LBS1} J.A.S. Lima, S. Basilakos, J. Sol\`a, Gen.Rel.Grav. {\bf 47} (2015) 40.

\bibitem{LBS2} J.A.S. Lima, S. Basilakos, J. Sol\`a,  Eur.Phys.J. C{\bf 76} (2016)  228.

\bibitem{LBS3}  J.A.S. Lima, S. Basilakos, J. Sol\`a,  Mon.Not.Roy.Astron.Soc. {\bf 431} (2013) 923; Int.J.Mod.Phys. D{\bf 22} (2013) 1342008; E.L.D. Perico et al, Phys.Rev. D{\bf 88} (2013) 063531.


\bibitem{Fossil07}  J. Sol\`a, J. Phys.  A\textbf{41} (2008) 164066.


\bibitem{SolaGomCruzApJL}  	
J. Sol\`a, A. G\'omez-Valent, J. de Cruz P\'erez, Astrophys. J. {\bf 811} (2015) L14.

\bibitem{SolaGomCruzApJ}  	
J. Sol\`a, A. G\'omez-Valent, J. de Cruz P\'erez,
Astrophys. J. {\bf 836} (2017) 43.

\bibitem{PRL2017}  	
J. Sol\`{a}, J. de Cruz P\'erez, A. G\'omez-Valent \& R.C. Nunes,  \textit{Dynamical Vacuum against a rigid Cosmological Constant}, arXiv:1606.00450.


\bibitem{SolaGomCruzMPLA} J. Sol\`a, A. G\'omez-Valent, J. de Cruz P\'erez, Mod. Phys. Lett. A{\bf 32} (2017) 1750054.

\bibitem{BPS09} S. Basilakos, M. Plionis, J. Sol\`a, Phys. Rev. {\bf D80} (2009) {3511}.

\bibitem{GSBP11}   J. Grande, J. Sol\`a, S. Basilakos, M. Plionis, \JCAP
    {\bf 1108} {2011} {007}.


\bibitem{FritzschSola2015} H. Fritzsch, J. Sol\`a, Mod. Phys. Lett. A{\bf 30} (2015) 1540034.

\bibitem{SolaKarim2017}
J. Sol\`a, E. Karimkhani, A. Khodam-Mohammadi,
Class. Quant. Grav. {\bf 34} (2017) 025006.

\bibitem{CCtPheno} M. Ozer and O. Taha, Phys. Lett. B{\bf 171} (1986) 363; Nucl. Phys. B{\bf 287} (1987) 776;
O. Bertolami, Nuovo Cimento B{\bf 93} (1986) 36;
K. Freese, F.C. Adams, J.A. Frieman, and E. Mottola, Nucl. Phys. B{\bf 287} (1987) 797; J. C. Carvalho, J. A. S. Lima and I. Waga, Phys. Rev. D{\bf 46} (1992) 2404.

\bibitem{Overduin} J. M. Overduin and F. I. Cooperstock, Phys. Rev. D{\bf 58} (1998) 043506.

\bibitem{Cristina2003} C. Espa\~na-Bonet et al., JCAP {\bf 0402} (2004) 006;
Phys.Lett. B{\bf 574} (2003) 149.

\bibitem{WangMeng2004}  P. Wang and X.-H. Meng,
Class. Quant. Grav. {\bf 22} (2005) 283.

\bibitem{AlcanizLima2005} J. S. Alcaniz, J. A. S. Lima,
Phys. Rev. D{\bf 72} (2005) 063516.

\bibitem{Hdata} X. L. Meng, X. Wang, S. Li, and T. J. Zhang, arXiv:1507.02517.

\bibitem{RatraFarooq2013}
O. Farooq \& B. Ratra, ApJ {\bf 766} (2013) L7.

\bibitem{RVMgrowth} S. Basilakos, J. Sol\`a,  Phys.Rev. D{\bf 92} (2015) 123501.

\bibitem{Ferreira}
M.C. Ferreira et al. Phys.Rev. D{\bf 89} (2014) 083011;
Phys.Rev. D{\bf 91} (2015) 124032.


\bibitem{alpha0} J.  Webb et al., Phys. Rev. Lett. \textbf{107}, 191101 (2011).

\bibitem{GUT} R. N. Mohapatra,
\textit{Unification and Supersymmetry: The Frontiers of Quark-Lepton Physics} (Springer, 2002).

\bibitem{CalmetFritzsch} X. Calmet, H. Fritzsch, Europhys. Lett. {\bf 76} (2006) 1064;
    Phys. Lett. B{\bf 540} (2002) 173;
    Eur. Phys. J C{\bf 24} (2002) 639.

\bibitem{alpha1} A. Songaila and L. Cowie, Astrophys. J. \textbf{793}, 103 (2014).


\bibitem{alpha2} T. M. Evans et al., MNRAS. \textbf{445}, 128 (2014).

\bibitem{alpha3} P.  Molaro et al.,Eur. Phys. J. ST \textbf{163} , 173 (2008).

\bibitem{alpha4} H. Chand et al.,  Astron. Astrophys. \textbf{451}, 45 (2006).

\bibitem{alpha6} P. Molaro et al., Astron. Astrophys. \textbf{555}, A68 (2013).

\bibitem{alpha5} I.  I.  Agafonova,  P.  Molaro,  S.  A.  Levshakov, and  J.  L.  Hou,  Astron. Astrophys. \textbf{529}, A28 (2011),



\bibitem{Rosenband2008} T. Rosenband et al.,
Science {\bf 319} (2008) 1808.



\bibitem{Bolotin2015} Yu.L. Bolotin, A. Kostenko, O.A. Lemets
\& D.A. Yerokhin, Int. J. Mod. Phys. D \textbf{24} (2015) 1530007.

 	
\bibitem{Salvatelli2014}
V. Salvatelli, N. Said, M. Bruni, A. Melchiorri \& D. Wands, Phys.Rev.Lett. {\bf 113} (2014) 181301.

 	
\bibitem{Murgia2016} 
R. Murgia, S. Gariazzo, N. Fornengo, JCAP {\bf 1604} (2016) 014.













\end{thebibliography}
\end{document}